\documentclass[manuscript]{aastex}

\usepackage{subfigure}
\usepackage{natbib}
\usepackage{epsfig}
\usepackage{graphicx}
\shorttitle{Dust evolution in the Taurus Complex}
\shortauthors{Flagey et al.}

\begin{document}
\DeclareGraphicsExtensions{.pdf,.gif,.jpg}

\title{Evidence for dust evolution within the Taurus Complex from Spitzer images}

\author{N. Flagey\altaffilmark{1,2},A. Noriega-Crespo\altaffilmark{1},F.Boulanger\altaffilmark{2},S.J. Carey\altaffilmark{1},T.Y. Brooke\altaffilmark{1},E. Falgarone\altaffilmark{3},T.L. Huard\altaffilmark{4},C.E. McCabe\altaffilmark{1,5},M.A. Miville-Desch\^enes\altaffilmark{2},D.L. Padgett\altaffilmark{1},R. Paladini\altaffilmark{1},L.M. Rebull\altaffilmark{1}}
\email{nflagey@ipac.caltech.edu}

\altaffiltext{1}{Spitzer Science Center, California Institute of Technology, 1200 East California Boulevard, MC 220-6, Pasadena, CA 91125, USA}
\altaffiltext{2}{Institut d'Astrophysique Spatiale, Universit\'e paris Sud, B\^at. 121, 91405 Orsay Cedex, France}
\altaffiltext{3}{LERMA/LRA, CNRS UMR 8112, \'Ecole Normale Sup\'erieure and Observatoire de Paris, 24 rue Lhomond, 75231 Paris Cedex 05, France}
\altaffiltext{4}{Smithsonian Astrophysical Observatory, 60 Garden Street, MS42, Cambridge, MA 02138, USA}
\altaffiltext{5}{Jet Propulsion Laboratory, MS 183-900, 4800 Oak Grove Drive, Pasadena, CA 91109, USA}

\begin{abstract}
We present Spitzer images of the Taurus Complex (TC). We take advantage of the sensitivity and the spatial resolution of the observations to characterize the diffuse infrared emission across the cloud. This work highlights evidence of dust evolution within the translucent sections of the archetype reference for studies of quiescent molecular clouds. We combine the Spitzer 160 $\mu$m and IRAS 100 $\mu$m observations to produce a dust temperature map and a far-IR dust opacity map at 5\arcmin\ resolution. The average dust temperature is about 14.5K with a dispersion of $\pm 1$K across the cloud. The far-IR dust opacity is tightly correlated with the extinction derived from 2MASS stellar colors and is a factor 2 larger than the average value for the diffuse ISM. This opacity increase and the attenuation of the radiation field both contribute to account for the lower emission temperature of the large grains. The structure of the TC significantly changes in the mid-IR images that trace emission from PAHs and very small grains (VSGs). We focus our analysis of the mid-IR emission to a range of ecliptic latitudes away from the zodiacal bands and where the zodiacal light residuals are small. Within this cloud area, there are no 8 and 24 $\mu$m counterparts to the brightest 160 $\mu$m emission features. Conversely, the 8 and 24 $\mu$m images reveal filamentary structure that is strikingly inconspicuous in the 160 $\mu$m and extinction maps. The infrared colors vary over sub-parsec distances across this filamentary structure. We compare the observed colors with model calculations quantifying the impact of the radiation field intensity and the abundance of stochastically heated particles on the dust SED. To match the range of observed colors, we have to invoke variations by a factor of a few of both the interstellar radiation field and the abundance of PAHs and VSGs. We conclude that within this filamentary structure a significant fraction of the dust mass cycles in and out the small size end of the dust size distribution.
\end{abstract}


\section{Introduction}

Understanding interstellar dust evolution is a major challenge underlying the interpretation of many infrared (IR) observations. The composition of interstellar dust is the outcome of processes in interstellar space that break and re-build grains over time scales much shorter than that of the dust renewal by stellar ejecta \citep{Draine1990, Tielens1998, Jones2005, Zhukovska2008}. There is a wide consensus on such a conclusion, but the processes that drive dust evolution in space are still poorly understood. Understanding the processes that control the small grain end of the dust size distribution is of particular importance because the smallest dust particles have a marked impact on the physics of the ISM. Polycyclic Aromatic Hydrocarbons (PAHs), and carbon nanometer Very Small Grains (VSGs) absorb a significant fraction of the UV starlight and are considered to be the dominant gas heating source in interstellar space penetrated by UV light. Given their large total surface area, it is also proposed that they act on gas depletion, chemistry, ionization degree and coupling with the magnetic field \citep[e.g.][]{Tielens2008}.

It is in molecular clouds where dust may grow by accretion of gas phase elements and grain-grain coagulation. Observations of the translucent sections ($A_V$ less than a few magnitudes) of molecular clouds are of interest to trace dust evolution that occurs before matters enter dense molecular cores and form stars. Many UV and IR studies provide evidence of diverse processes that impact the dust size distribution. The systematic change in the visible-to-UV extinction curve correlated with the $R_V = A_V /E(B - V)$ ratio \citep{Cardelli1989} has been interpreted as evidence for grain growth \citep{Kim1994}. Variations of the dust IR emission colors in translucent molecular clouds have been interpreted as evidence for changes in the small grains abundances and large grains emissivities \citep[e.g.][]{Bernard1993,Bernard1999,DelBurgo2003,Schnee2005,Kiss2006}. \citet{Stepnik2003} have shown that the deficit of small grains and enhancement of large grain emissivity within a quiescent dense filament can be interpreted as grain-grain coagulation. However, some UV observations do not fit with a simple evolutionary picture where the smallest grains progressively stick on the largest \citep{Boulanger1994, Mathis1994, Whittet2004}.

Correlations with density and velocity structure of clouds have also been measured. For example, \citet{Miville2002} observed an enhancement of PAHs abundance in a filament of Ursa Major characterized by a large transverse velocity gradient. \citet{Bernard1999} measured a drop in the PAHs abundance between H~I gas and denser molecular gas traced by CO emission. Mid-IR spectroscopic variations of photon-dominated regions suggest that photo-chemical processing of VSGs gives birth to free PAHs in PDRs \citep{Rapacioli2005,Berne2007,Compiegne2008}.

Most of the earlier work focuses on small sections of molecular clouds or is based on low resolution data \citep[e.g.][]{Lagache1998,Cambresy2005}. In this paper, we take advantage of the coverage, sensitivity and resolution of recent \textit{Spitzer} observations to present a global study of the nearby Taurus Complex (TC), a giant molecular complex about 140 parsecs from us \citep{Kenyon1994} characterized by a lack of young massive stars and clusters. The TC is perfectly suited for high spatial resolution studies, mainly focused on low mass star formation \citep[e.g.][]{Hartmann2002, Palla2002, Nakamura2008} and interstellar chemistry \citep[e.g. ][]{Pratap1997, Wolkovitch1997, Maezawa1999, Sunada1999, Harju2000, Whittet2007, Goldsmith2008}. The TC also is the ideal region to analyze the structure of the interstellar medium (ISM) since it corresponds to the intermediary phase between diffuse, turbulent, non structured medium and the massive star forming regions dominated by the radiation field of young OB stars. However, only a few global studies have been performed through infrared (IR) observations. \citet{Langer1989} and \citet{Abergel1994} have shown that the far-IR (FIR) emission as seen by IRAS was correlated to $^{13}$CO emission and extinction, and that the cold dust component (low 60 $\mu$m to 100 $\mu$m emission ratio) was particularly well traced by the gas. \citet{Stepnik2003} have observed significant variations of the IR colors within a TC filament which may be related to an evolution of the dust properties as one enters or exits the most extinguished regions of the cloud. Analysis of extinction within the TC by \citet{Whittet2004} and \citet{Shenoy2008} have also highlighted variations of dust properties.

We present the \textit{Spitzer} observations of the cloud and the way we process them to build complete maps of the extended emission in section \ref{lab:obs}, focusing on the zodiacal light subtraction issue. Section \ref{lab:res} presents our results on the average dust emission spectrum, the dust temperature variations and the correlation between FIR emission, dust temperature and visual extinction. We analyze a small sub-area of the TC in section \ref{lab:local} to further constrain the local variations of the dust emission colors and relate them to variations of the ISRF intensity and dust size distribution in section \ref{lab:disc}.

\section{Observations}
\label{lab:obs}

\subsection{Spitzer observations}

\begin{figure*}[!t]
\centering
\subfigure[]
	{\label{}
	\includegraphics[angle=90,width=.475\linewidth]{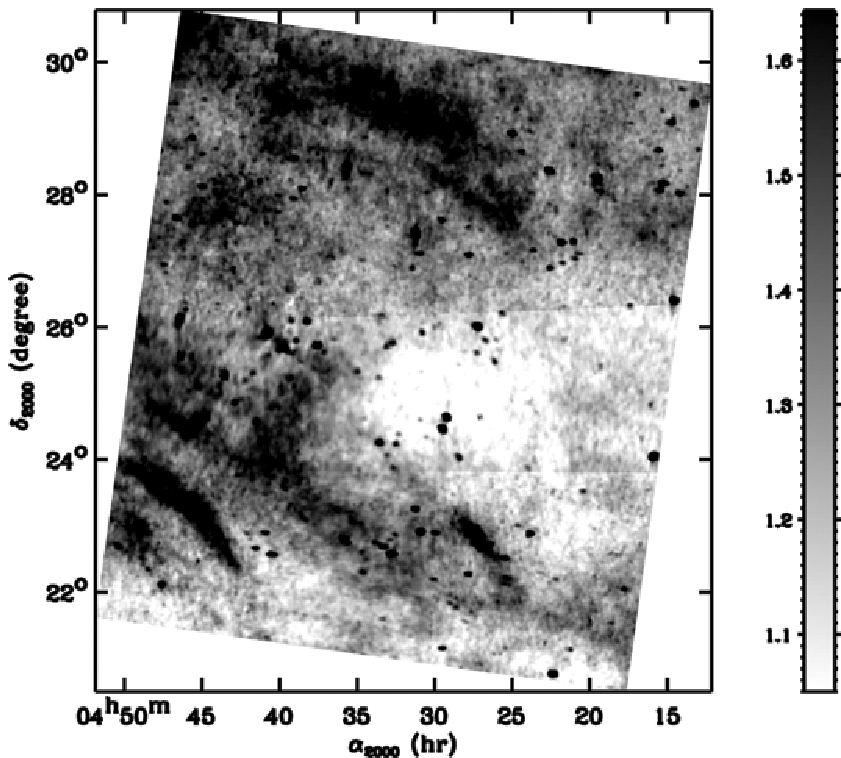}}
\hspace{.0125\linewidth}
\subfigure[]
	{\label{}
	\includegraphics[angle=90,width=.475\linewidth]{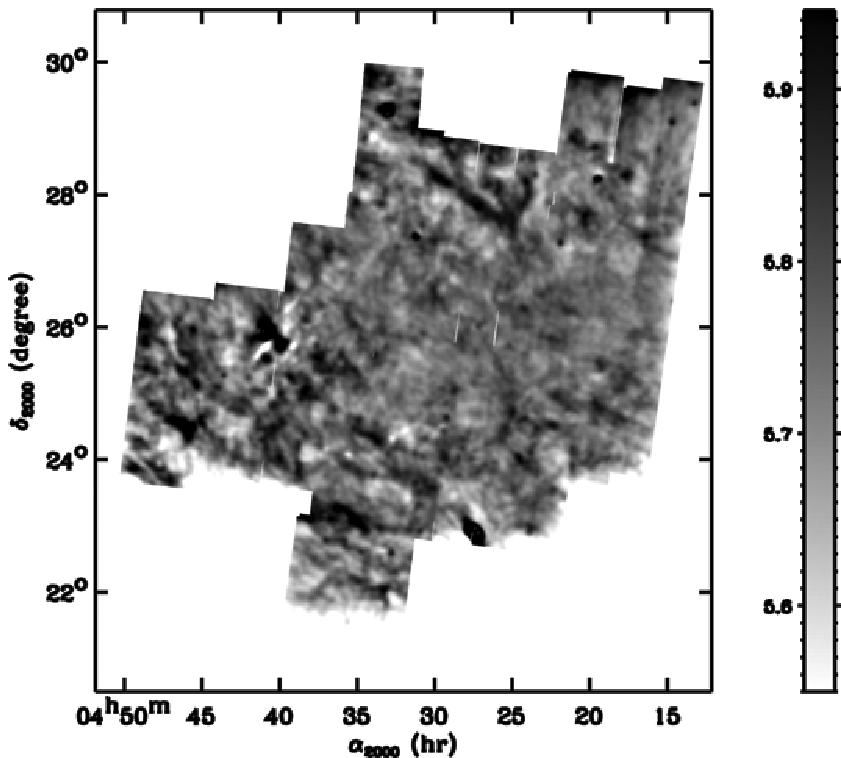}}
\caption{MIR emission of the TC as seen by (a) IRAS 12 and (b) IRAC 8 $\mu$m at the resolution of IRAS 100 (4.3\arcmin) and after point source subtraction. The units are in MJy/sr.}
\label{fig:data_12_8}
\end{figure*}

\begin{figure*}[!t]
\centering
\subfigure[]
	{\label{}
	\includegraphics[angle=90,width=.475\linewidth]{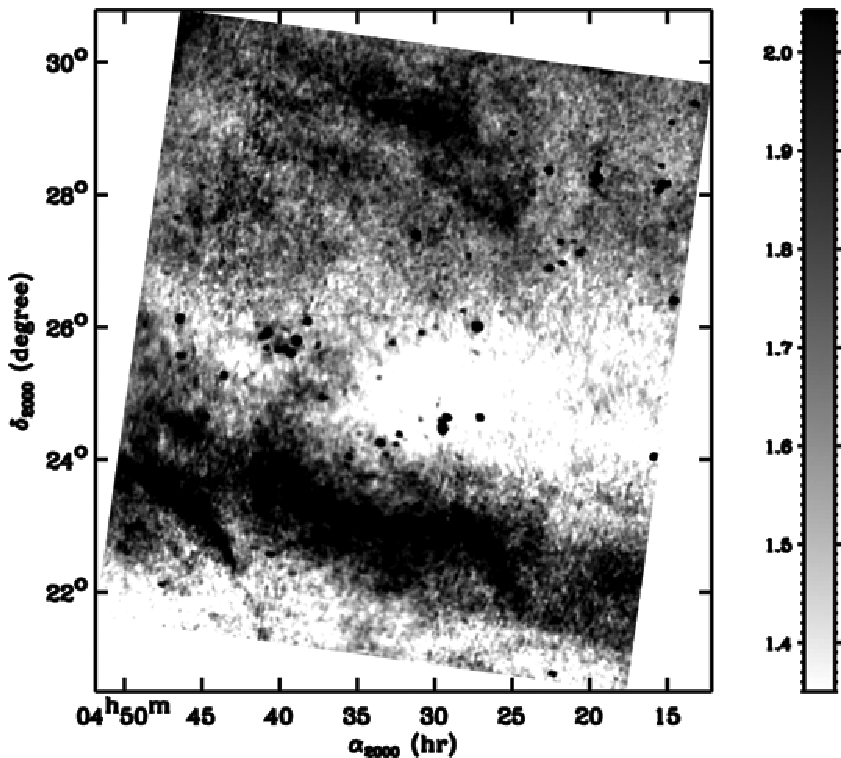}}
\hspace{.0125\linewidth}
\subfigure[]
	{\label{}
	\includegraphics[angle=90,width=.475\linewidth]{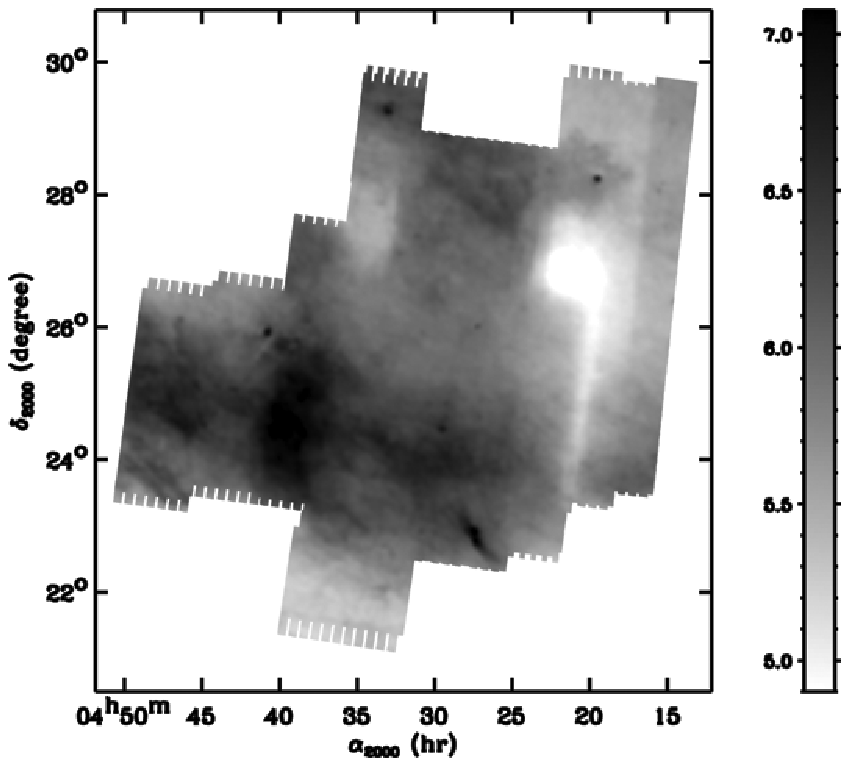}}
\caption{MIR emission of the TC as seen by (a) IRAS 25 and (b) MIPS 24 $\mu$m at the resolution of IRAS 100 (4.3\arcmin) and after point source subtraction. A zodiacal band appears clearly on both images at a different position due to the change in parallax. On the MIPS 24 $\mu$m images, offsets due to the combination of different epochs observations are also visible. The units are in MJy/sr.}
\label{fig:data_25_24}
\end{figure*}

\begin{figure*}[!t]
\centering
\subfigure[]
	{\label{}
	\includegraphics[angle=90,width=.475\linewidth]{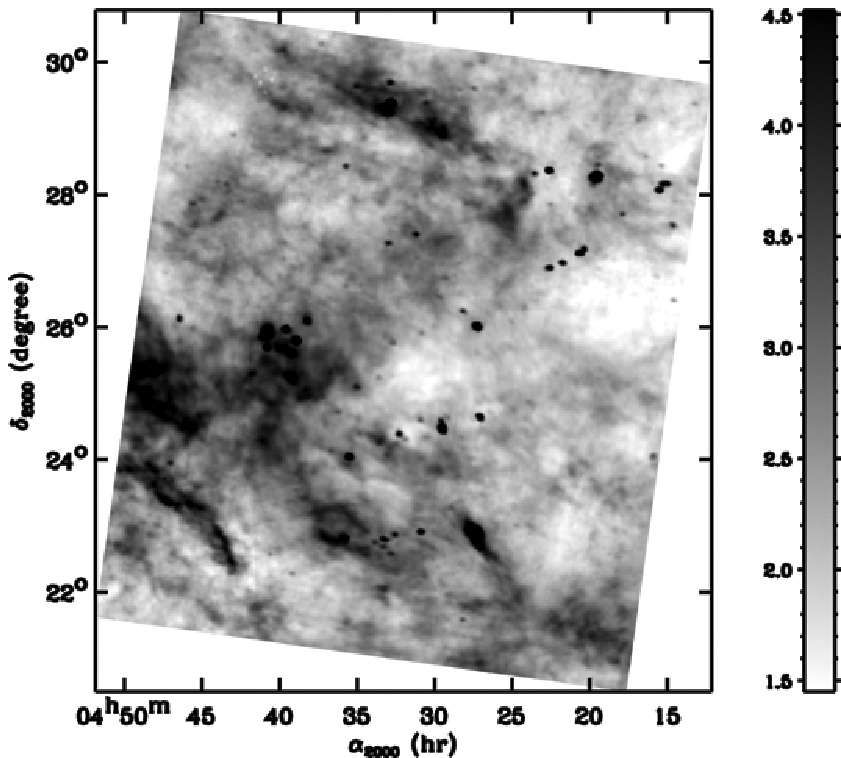}}
\hspace{.0125\linewidth}
\subfigure[]
	{\label{}
	\includegraphics[angle=90,width=.475\linewidth]{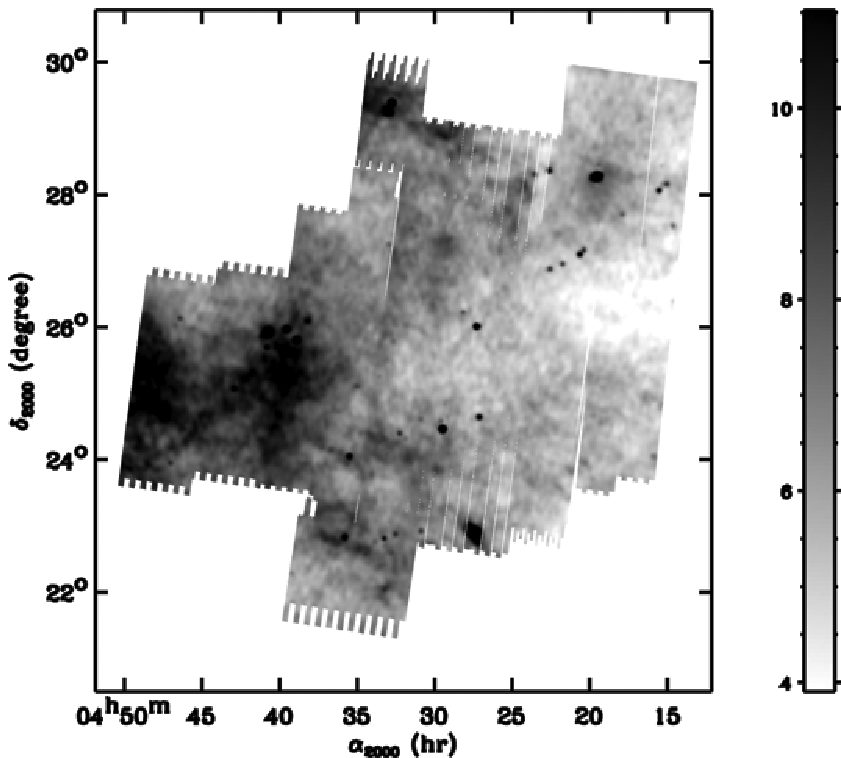}}
\caption{FIR emission of the TC as seen by (a) IRAS 60 and (b) MIPS 70 $\mu$m at the resolution of IRAS 100 (4.3\arcmin). The units are in MJy/sr.}
\label{fig:data_60_70}
\end{figure*}

\begin{figure*}[!t]
\centering
\subfigure[]
	{\label{}
	\includegraphics[angle=90,width=.475\linewidth]{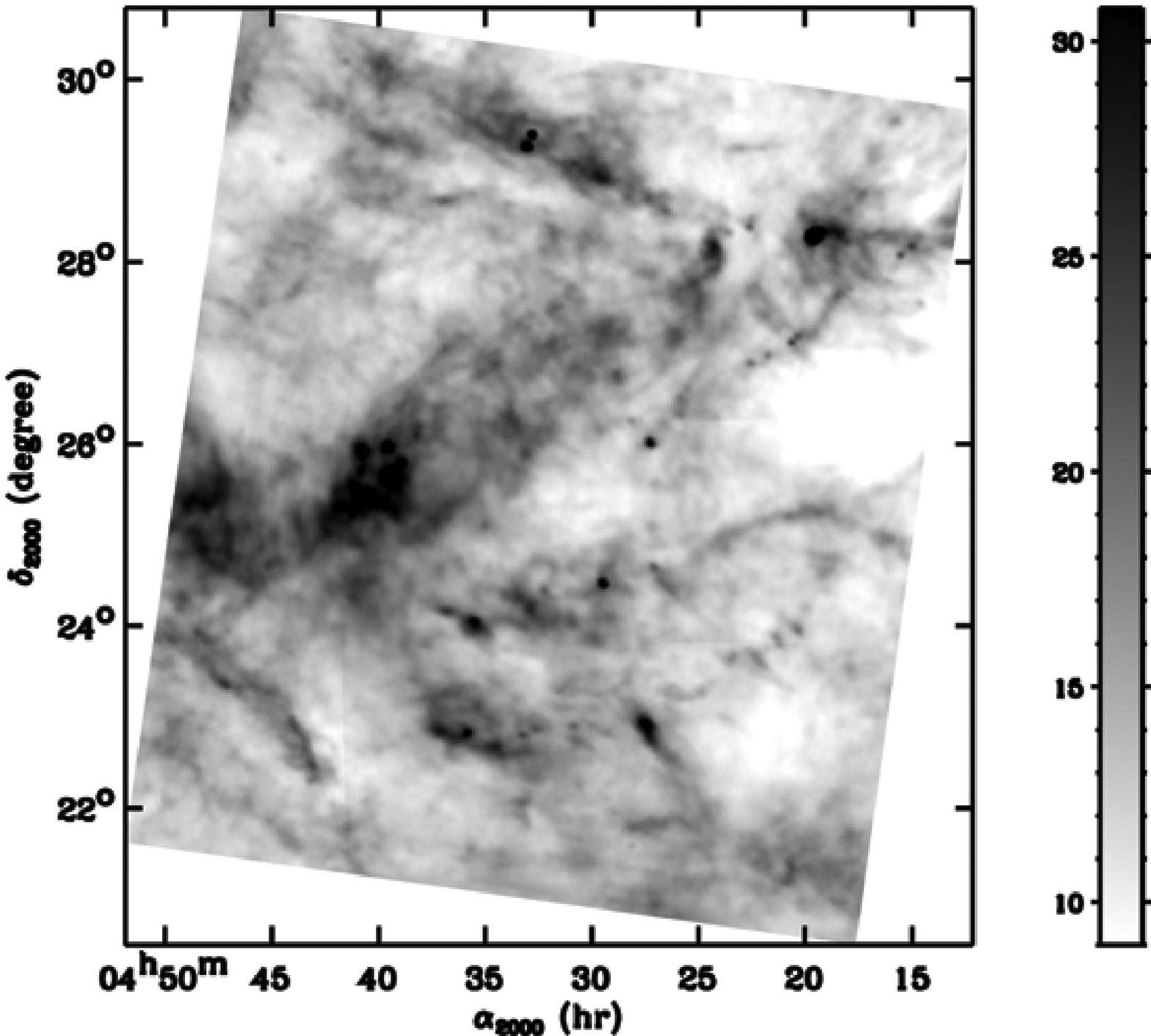}}
\hspace{.0125\linewidth}
\subfigure[]
	{\label{}
	\includegraphics[angle=90,width=.475\linewidth]{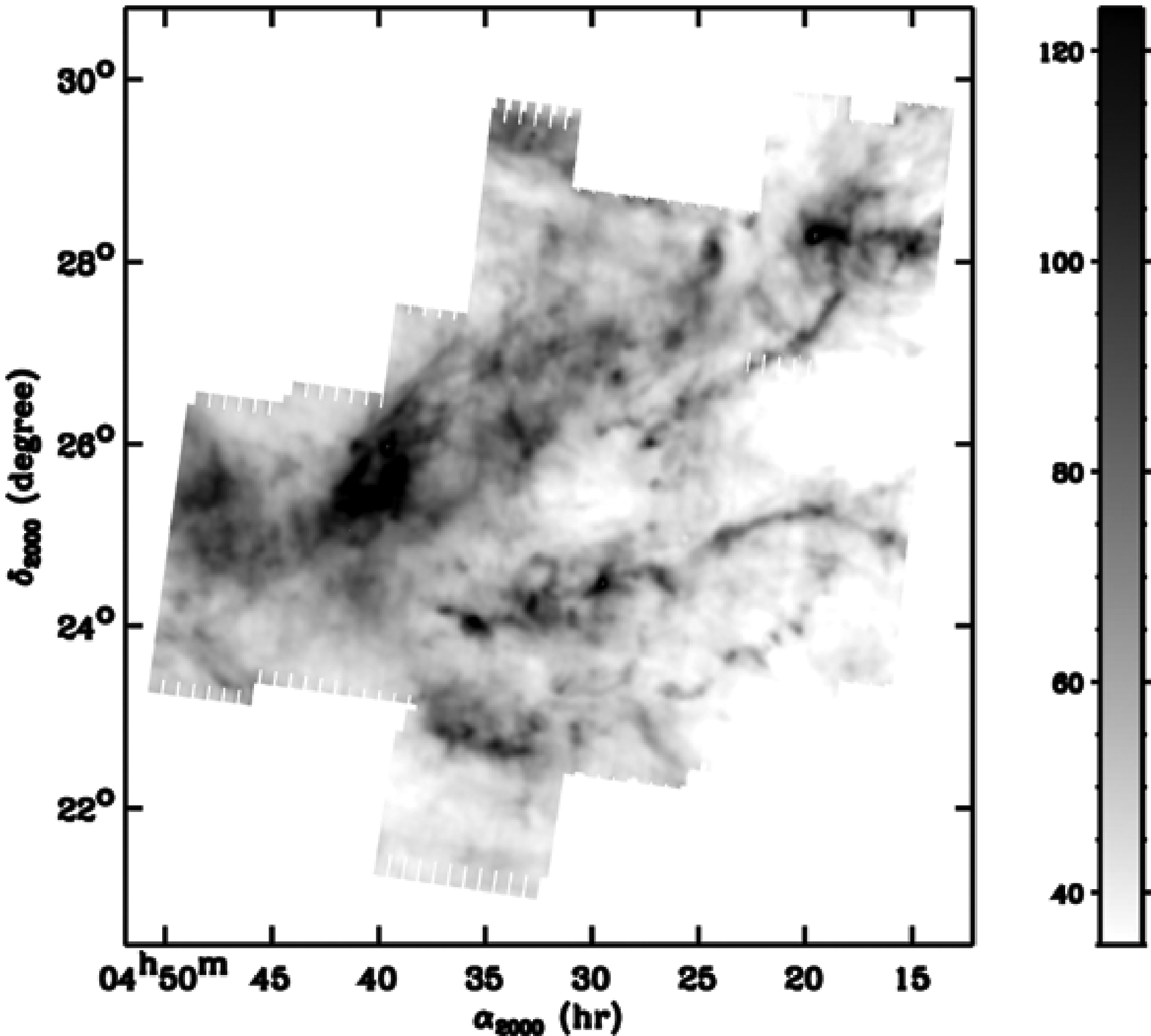}}
\caption{FIR emission of the TC as seen by (a) IRAS 100 and (b) MIPS 160 $\mu$m at the resolution of IRAS 100 (4.3\arcmin). The units are in MJy/sr.}
\label{fig:data_100_160}
\end{figure*}

The bulk of our analysis relies on recent data obtained by \textit{Spitzer} using the IRAC and MIPS instruments and are part of the Taurus Legacy project \citep{Padgett2009}. Briefly, the data come from three Spitzer programs (PIDs 3584, 30816 \& 462), use two epochs to find and weed out asteroids, and cover a total area of approximately 44 square degrees. The IRAC data in their four bands (3.6, 4.5, 5.8 \& 8 $\mu$m) utilize the 12 seconds integration High Dynamic Range (HDR) frame mode \citep{Fazio2004} with 2 dither positions per epoch. In the present study, we only use the IRAC 8 $\mu$m channel. The MIPS data in their three bands (24, 70 \& 160 $\mu$m) utilize the fast scan mode with a 3 seconds integration time per frame and 5 visits per pixel. Because of the different size of the three arrays \citep{Rieke2004} only the 24 $\mu$m maps have full coverage twice (two epochs). The 70 and 160 $\mu$m maps do cover the entire survey region, but only to a depth of 15 seconds integration per pixel. We use the three MIPS channels in the present study.

As part of the Legacy commitment to improve the data, most of the obvious artifacts (latencies, stray light, cosmic rays, etc) for both IRAC and MIPS are removed in the individual frames (the Basic Calibration Data or BCD), before the creation of the final mosaics. Two additional steps are carried out also at the BCD level prior to mosaicking: zodiacal light removal and overlap correction. The first step uses the \textit{Spitzer Science Center} (SSC) zodiacal light model based on \citet{Kelsall1998}, while the second step matches (globally) the median sky level across frames by applying a small offset (either positive or negative). Although these two steps rely on standard techniques and our best knowledge of the sky at mid/far infrared wavelengths, we do check processing does not modify the overall levels of the diffuse emission. Therefore, we compare our mosaics to the IRIS data \citep[reprocessed IRAS data by][]{Miville2005b} at 12, 25, 60 and 100$\mu$m (see Fig~.\ref{fig:data_12_8}, \ref{fig:data_25_24}, \ref{fig:data_60_70} and  \ref{fig:data_100_160}).

\subsection{Zodiacal light residuals}

The TC is located at a low ecliptic latitude ($\beta < 10\degr$). We use the model of \citet{Kelsall1998} to estimate the zodiacal light surface brightness. They significantly change in the multi-epoch Spitzer data. We only report the values of the first epoch: $\sim$10 MJy/sr at 8 $\mu$m, $\sim$50 MJy/sr at 24 $\mu$m, $\sim$15 MJy/sr at 70 $\mu$m and $\sim$5 MJy/sr at 160 $\mu$m. The total surface brightness observed by \textit{Spitzer} are about 15, 55, 25 and 135 MJy/sr at 8, 24, 70 and 160 $\mu$m respectively. The contribution of the zodiacal light to the infrared emission is significant at all wavelengths shortward of 160 $\mu$m. The subtraction of this foreground component is thus important but difficult, as discussed below.

Subtraction of the zodiacal light component for the IRAC data is done as follows. Each 5 by 5 arcmin IRAC BCD image has an estimate of the zodiacal light associated to its header (ZODY\_EST keyword). This estimate is also based on the model of \citet{Kelsall1998}. However, the \textit{Spitzer} pipeline subtracts from each BCD the contribution of the sky (including zodiacal light) at the sky bias observation position (SKYDKRZB keyword). In order to fully subtract the zodiacal light in agreement with the \citet{Kelsall1998} model, we have to subtract the difference (ZODY\_EST - SKYDKRZB) for each single BCD when building the mosaic. A correction factor for extended emission is also required, as given in the IRAC Data Handbook\footnote{See http://ssc.spitzer.caltech.edu/irac/dh}. Comparison with IRIS 12 $\mu$m images shows that some structures have not moved in more than 20 years (see Fig.~\ref{fig:data_12_8}). We thus believe these structures are not associated with zodiacal bands \citep{Low1984}, which are not taken into account by the model of \citet{Kelsall1998}.

For MIPS 24 and 70 $\mu$m data the steps described above for IRAC are not necessary since the image bias is estimated a different way. The mosaic is built with a correction that takes into account the predicted value of the zodiacal light for each BCD. As for IRAC 8 $\mu$m, we compare MIPS 24 and 70 $\mu$m to IRIS 25 and 60 $\mu$m images (see Fig.~\ref{fig:data_25_24} and \ref{fig:data_60_70}). At 24 and 25 $\mu$m, there are some structures that have not moved between the time they have been observed by IRAS and by \textit{Spitzer}. However, it appears on the MIPS 24 $\mu$m mosaic that there is an emission component of constant ecliptic latitude $2\degr < \beta < 3\degr$. This component is also visible on IRIS 25 $\mu$m images, but at a slightly different position, for $1\degr < \beta < 2\degr$ (see Fig.~\ref{fig:data_25_24}). The positional difference of the features between epochs suggests that it is a zodiacal dust band whose position on the sky has changed between IRAS and \textit{Spitzer} observations due to parallax. On the MIPS 24 $\mu$m mosaic, there are also some offsets that appear between the different epochs when the TC observations have been scheduled. We assume these offsets come from the uncertainty of the zodiacal light model. At 60 and 70 $\mu$m, the emission structure is almost the same and we are confident that the contribution of the zodiacal light residuals, including bands, is negligible.

We estimate the residuals associated with the zodiacal band by median filtering the MIPS 24 $\mu$m mosaic as a function of the ecliptic latitude. The standard deviation of the resulting profile is about 0.3 MJy/sr. We compute the zodiacal light IR colors within the TC at the first two epochs of the observations thanks to the model of \citet{Kelsall1998}. We use them to scale the MIPS 24 $\mu$m measurement to the other wavelengths. We obtain standard deviations of less than 0.03 MJy/sr for MIPS 160 $\mu$m, about 0.1 MJy/sr for MIPS 70 $\mu$m and up to 0.5 MJy/sr for IRAC 8 $\mu$m. The spatial variations of the residual maps are about 21, 3.4, 0.47 and 0.15 MJy/sr at 160, 70, 24 and 8 $\mu$m respectively. The contribution of the zodiacal band is not significant at MIPS 160 $\mu$m or at MIPS 70 $\mu$m. However, it is an important component of the emission variations for MIPS 24 $\mu$m and IRAC 8 $\mu$m. Thus, while we will conduct the analysis of the extended emission across the entire TC at wavelengths strictly longer than 24 $\mu$m, we will adapt our analysis at shorter wavelengths in such a way we avoid the main zodiacal band.

\subsection{Point source subtraction}

We are interested in the extended emission across the TC. Therefore, we have to remove the point sources, which are mostly present at shorter wavelengths (IRAC 8 $\mu$m and MIPS 24 $\mu$m). We first produce a filtered map resulting from a median filter of the data by a box of 3\arcmin x3\arcmin. We subtract this median filtered image from the original mosaics. Finally, every pixel in the difference which is more than $3\sigma$ away from the median value within a box of 3\arcmin x3\arcmin is flagged and replaced by interpolated value of neighboring pixels.

The source removed \textit{Spitzer} data at 8, 24, 70 and 160 $\mu$m are shown at the resolution of IRAS 100 $\mu$m (4.3\arcmin) observations on Figures \ref{fig:data_12_8}, \ref{fig:data_25_24}, \ref{fig:data_60_70} and \ref{fig:data_100_160}. For sake of comparison, IRIS images at 12, 25, 60 and 100 $\mu$m are also shown.

\section{Data analysis}
\label{lab:res}

We begin our analysis by comparing the infrared observations with each other on a global scale. Some of these observations can be related to one dust component, especially for \textit{Spitzer} observations: IRAC 8 $\mu$m traces PAHs, MIPS 24 $\mu$m traces VSGs and MIPS 160 $\mu$m traces BGs. The comparison between the shorter, intermediate and longer wavelengths observations provide us with informations regarding the variations of the PAHS, VSGs and BGs relative abundances, their properties and/or their excitation conditions. We combine \textit{Spitzer}, IRIS as well as the Zodi-Subtracted Mission Average Maps of DIRBE to build an average spectral energy distribution (SED) of the TC and compare it to a diffuse medium reference \citep{Flagey2007}. We use this average SED to constrain average dust properties across the TC. We then analyze of spatial variations by building a map of the big grains temperature and FIR emissivity that is compared to FIR emission and visual extinction.

\subsection{Global comparison of infrared observations from 8 to 240 $\mu$m}

We examine the available infrared observations from 8 to 240 $\mu$m: IRAC 8 $\mu$m, MIPS 24 $\mu$m, IRIS and DIRBE 60 $\mu$m, MIPS 70 $\mu$m, IRIS and DIRBE 100 $\mu$m, DIRBE 140 $\mu$m, MIPS 160 $\mu$m and DIRBE 240 $\mu$m. The data (except DIRBE channels) are shown at the resolution of IRAS 100 $\mu$m observations (4.3\arcmin) on Figures \ref{fig:data_12_8} to \ref{fig:data_100_160}. The emission morphology is significantly different between shorter and longer wavelengths with the 60, 70, 100 and 160 $\mu$m data showing a different correlation and the 8 and 24 $\mu$m emission exhibiting a strong correlation. It thus appears that, away from the regions highly contaminated by the zodiacal bands, emission from PAHs and VSGs are well correlated with each other while they are not with that of BGs as further discussed in Section \ref{lab:local}.

\subsection{Global dust spectral energy distribution}
\label{lab:sed}

To build the SED, we plot the flux at each wavelength as a function of the immediate longer wavelength (e.g. MIPS 70 $\mu$m versus IRIS 100 $\mu$m). Whenever possible, we measure the average color of the dust emission by fitting this pixel-to-pixel correlation by a straight line. Three separate IRIS plates cover the entire TC and the correlations are done on a plate by plate basis to avoid offset issues \citep{Miville2005b}. In order to constrain efficiently the dust properties, we extend the SED as far as possible towards the longer wavelengths and add the DIRBE colors to \textit{Spitzer} and IRIS measurements.

\begin{figure}[!t]
	\centering
	\includegraphics[angle=90.,width=.75\linewidth]{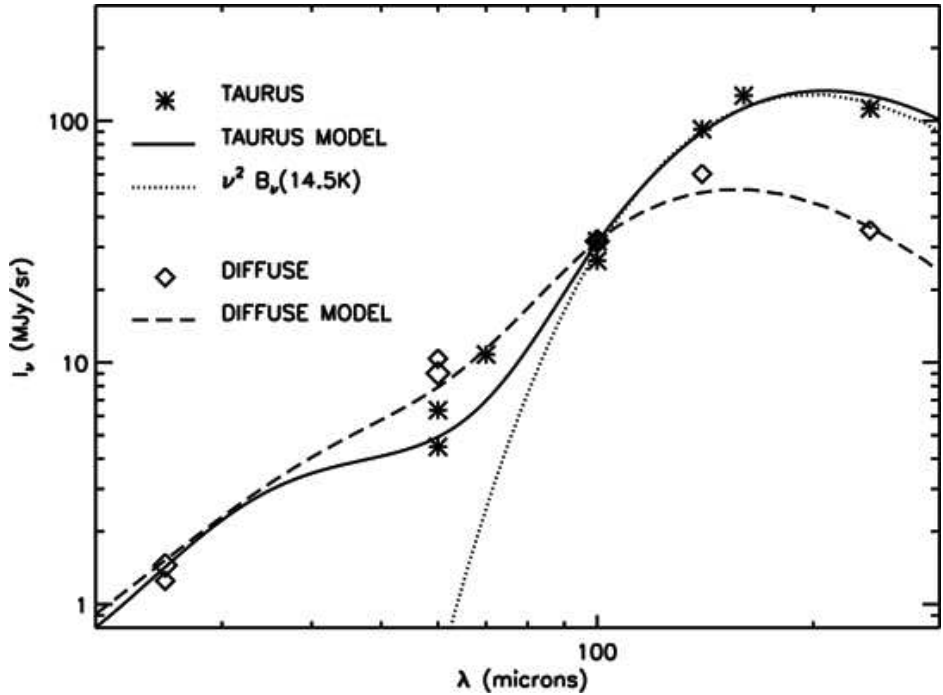}
	\caption{Far infrared mean spectral energy distribution of the TC (\textit{stars}) and the dust model that fits it (\textit{solid line}) as well as the $\nu^2 B_\nu(T)$ function that best fits the FIR section of the SED. For the sake of comparison, we show the SED of the extended mean Galactic diffuse emission (\textit{diamonds}) and the dust model that fits it (\textit{dashed}). The TC average SED is normalized at the DIRBE 100 $\mu$m diffuse brightness.}
	\label{fig:seds}
\end{figure}

Each plate exhibits a strong correlation between IRIS 60 $\mu$m and MIPS 70 $\mu$m, MIPS 70 $\mu$m and IRIS 100 $\mu$m, IRIS 100 $\mu$m and MIPS 160 $\mu$m, which are all dominated by the BG component. The average dust emission colors are MIPS70/IRIS60 = $2.2 \pm 0.3$, IRIS100/MIPS70 = $2.9 \pm 0.3$, MIPS160/IRIS100 = $4.0 \pm 0.15$. The uncertainties on the TC colors take into account the variations from one IRIS plate to another. We also measure, at the resolution of DIRBE, the DIRBE60/DIRBE100, DIRBE100/DIRBE140 and DIRBE140/DIRBE240 colors as well as the DIRBE100/IRIS100 color to match DIRBE colors with the previous measurements (see Tab.~\ref{tab:dirbe}). We show the average FIR SED of the TC in Figure \ref{fig:seds} along with that of the diffuse medium from \citet{Flagey2007}.

\subsection{Dust temperature}
\label{lab:dusttemp}

\begin{figure}[!t]
	\centering
	\includegraphics[angle=90.,width=.75\linewidth]{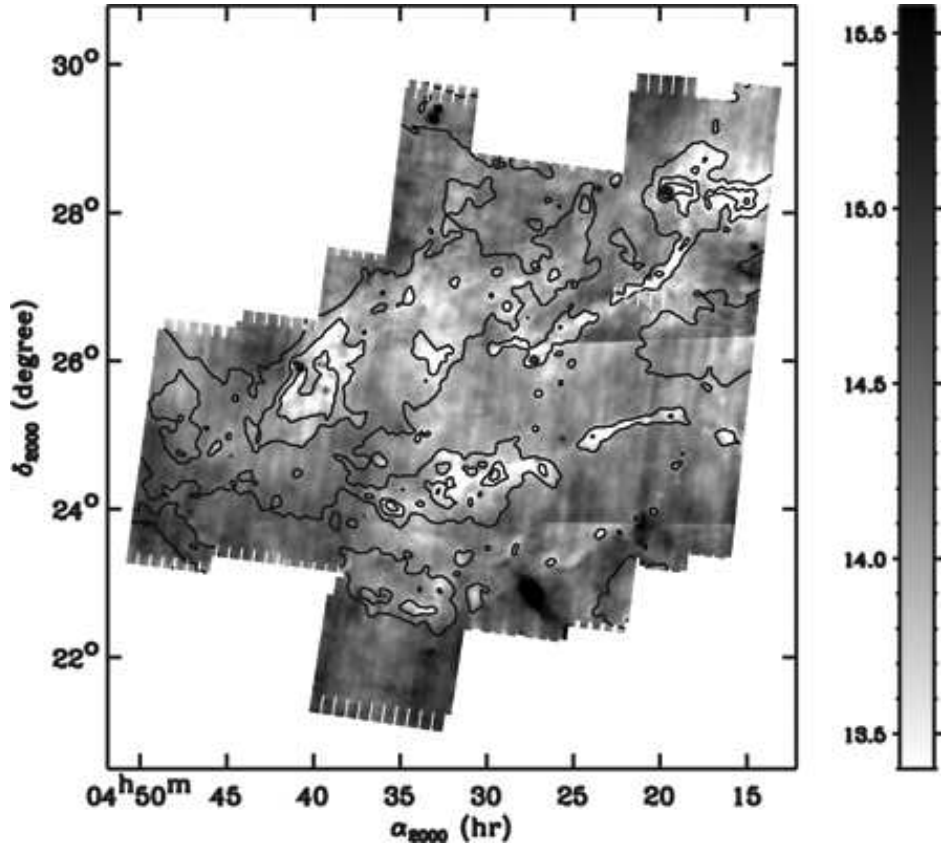}
	\caption{Dust temperature map as deduced from the IRIS 100 $\mu$m to MIPS 160 $\mu$m brightnesses ratio and a $\nu^2 B_\nu(T)$ function. The contours are those from the MIPS 160 $\mu$m map smoothed to the IRAS 100 $\mu$m resolution of 4.3\arcmin. The slight level jumps in the map (e.g. at a declination of $\sim 26.5\degr$) are due to the offsets between IRIS plates.}
	\label{fig:dusttempmap}
\end{figure}

\begin{figure}[!t]
	\centering
	\includegraphics[angle=90.,width=.75\linewidth]{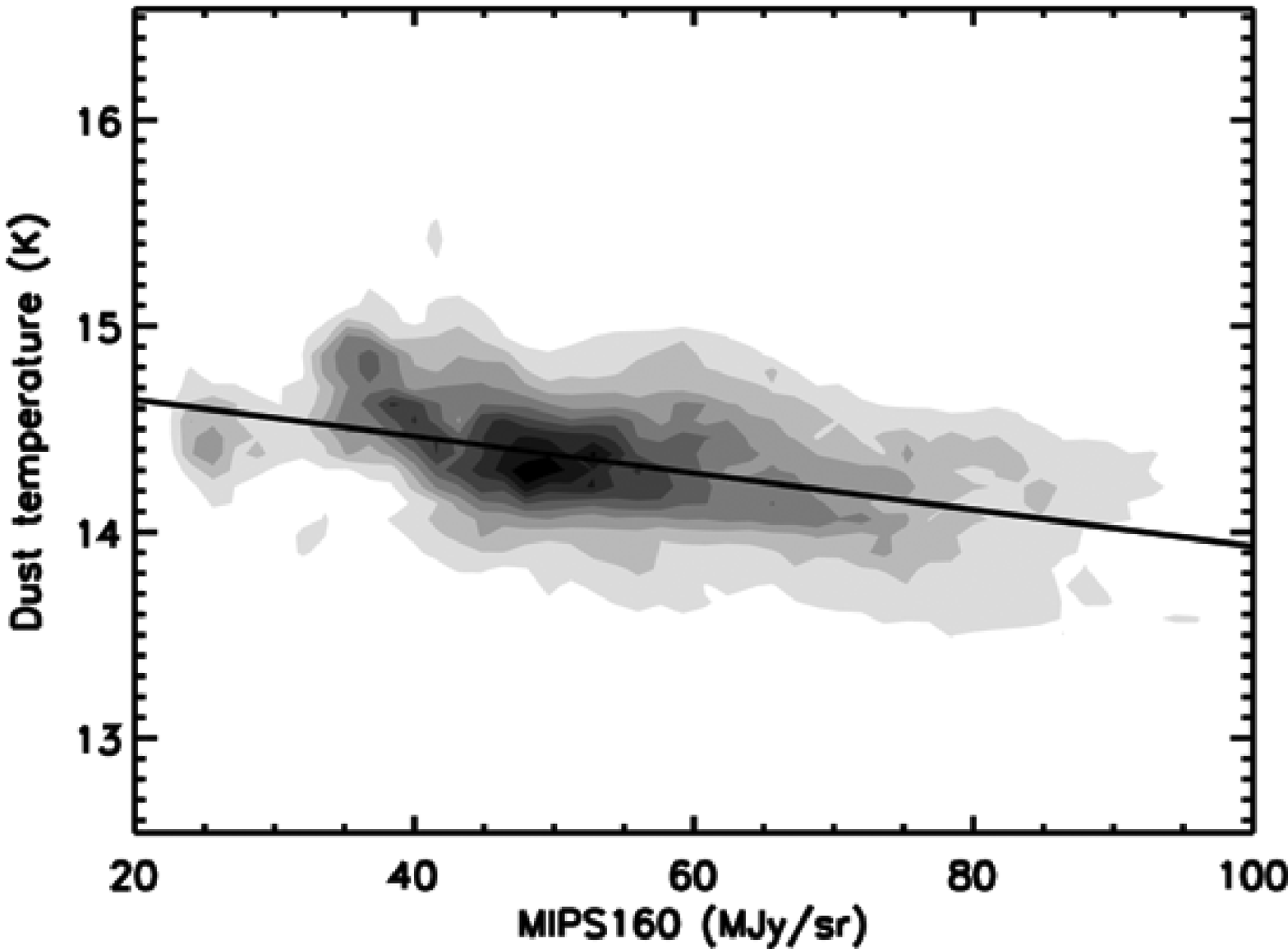}
	\caption{Correlation plot of the dust temperature as deduced from IRIS 100 and MIPS 160 $\mu$m observations as a function of MIPS 160 $\mu$m brightness. The plot corresponds to one IRIS plate only.}
	\label{fig:dusttempplot}
\end{figure}

We use the average SED of the TC to constrain the BGs temperature, focusing on the wavelengths from 100 to 240 $\mu$m, where this component is expected to dominate the emission. We fit this reduced SED by a $\nu^\beta B_\nu(T)$ function. We assume $\beta=2$, which is the value observed in the diffuse ISM \citep{Boulanger1996}. It is also that of the large grains in the \citet{Desert1990} model and of graphite and silicate grains in the \citet{Draine2007} model. The best fit, shown on Figure \ref{fig:seds} as a dotted line, corresponds to an average BGs temperature of $14.5\pm0.2\ \rm{K}$. This value is in agreement with previous measurement of the BGs temperature within a filament of the TC by \citet{Stepnik2003}. They have measured a large scale temperature of 16.8 K and an envelope temperature of 14.8 K using IRAS 100 $\mu$m and PRONAOS 580 $\mu$m observations \citep{Ristorcelli1998}. They also noted a drop of the 60 $\mu$m emission relative to the 100 $\mu$m that they explained by a depletion of the VSGs and by the introduction of a cold BGs component in addition to the warmer BG component. The cold component has a temperature of 12 K. \citet{delBurgo2005} used ISOPHOT and IRAS measurements from 60 to 200 $\mu$m to separate the dust emission into a \textit{cold} and a \textit{warm} component for the Taurus Molecular Cloud TMC-2 and its surrounding. Their temperatures are 12.5 and 20 K respectively. Combining IRAS and Spitzer observations of the Perseus molecular cloud, \citet{Schnee2008} build a dust temperature map that shows the same range of values as within the TC. Such decomposition cannot be detected in our analysis as we may average it with the warmer envelope. We thus neither confirm nor question such a result at this time. Instead, we examine variations of dust temperature across the TC.

The pixelwise dispersion in the observed emission ratios for the longer wavelengths may be accounted for by local variations of the dust properties and more particularly the BGs temperature. In order to analyze its spatial variations down to the smallest possible angular scale, we build a map of the BGs temperature across the entire cloud using the temperature that best fits the ratio between the fluxes at 100 and 160 $\mu$m. The resulting map is shown on Figure \ref{fig:dusttempmap}. The mean value of the BGs temperature is 14.5 K and the spatial dispersion through the entire cloud is about $\pm$1 K. With only two FIR data points at 100 and 160 $\mu$m we can not constrain the spatial variations of $\beta$ as well and thus keep $\beta=2$ throughout this process. Had we assumed a lower value of $\beta$, the absolute values of the temperature map would be larger but its structure would be the same.
 
The BGs temperature dispersion appears to be related to the structure of the cloud. Cold filaments seem to be in the middle of a warmer medium. The temperature map is well correlated with the dust emission in the FIR. The cold structures are probed by the high fluxes at 160 $\mu$m, as shown on Figure \ref{fig:dusttempplot}. However, the FIR dust temperature we are using is an effective temperature. It corresponds to a line-of-sight average where the value at each depth into the cloud is weighted by the local emission. Since the FIR dust emission steeply increases with dust temperature, the effective temperature is biased towards the largest temperatures, those at the surface of the clouds. Smaller values of the dust temperature ($\lesssim 12$ K) have been derived from sub-millimiter observations of dark clouds within the TC \citep{Stepnik2003, Schnee2005, Schnee2007}. This is probably due to the longer wavelength nature of the data.

\subsection{Dust opacity and extinction maps}

\begin{figure*}[!t]
\centering
\subfigure[]
	{\label{fig:av}
	\includegraphics[angle=90.,width=.475\linewidth]{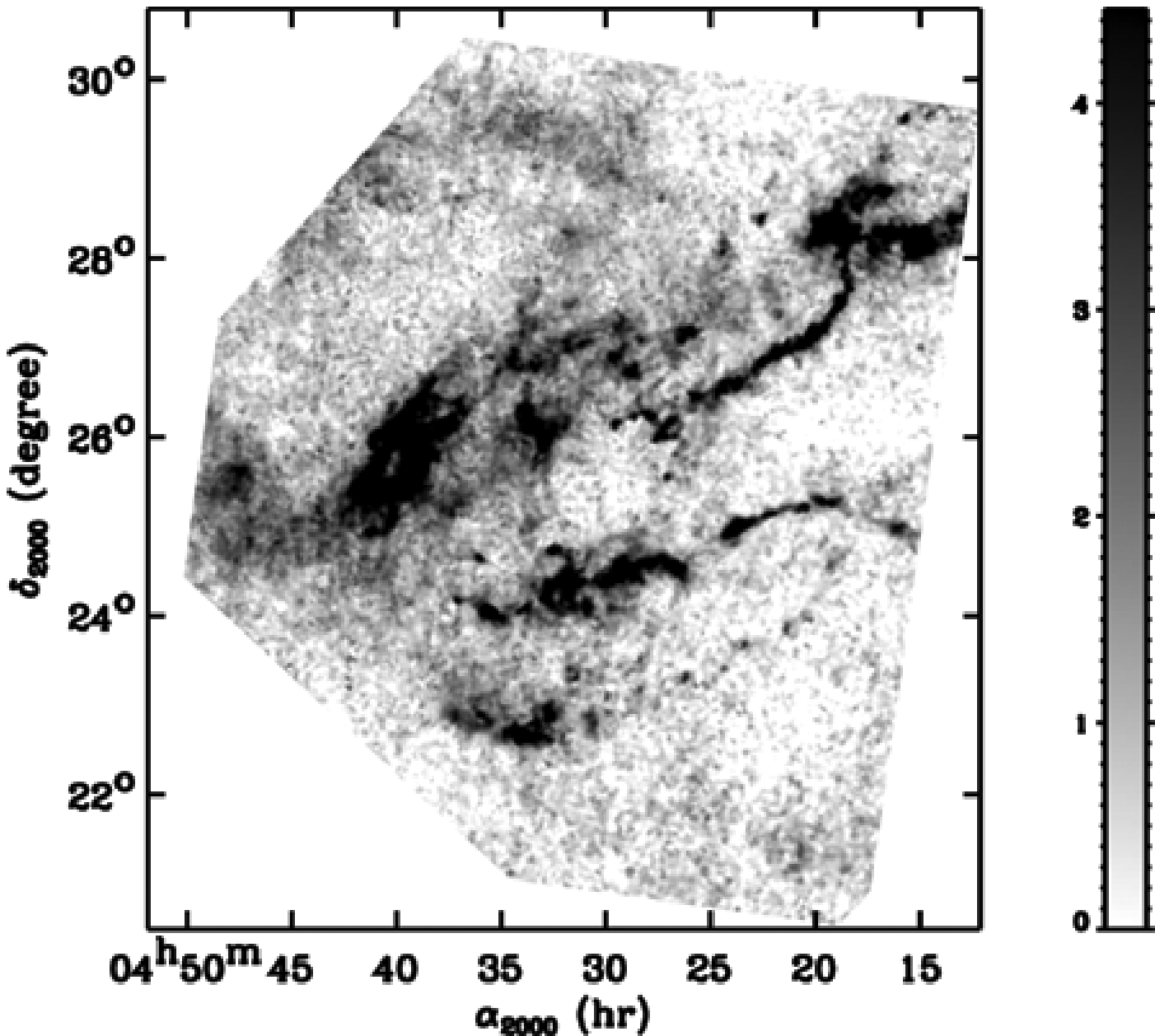}}
\hspace{.0125\linewidth}
\subfigure[]
	{\label{fig:taumap}
	\includegraphics[angle=90.,width=.475\linewidth]{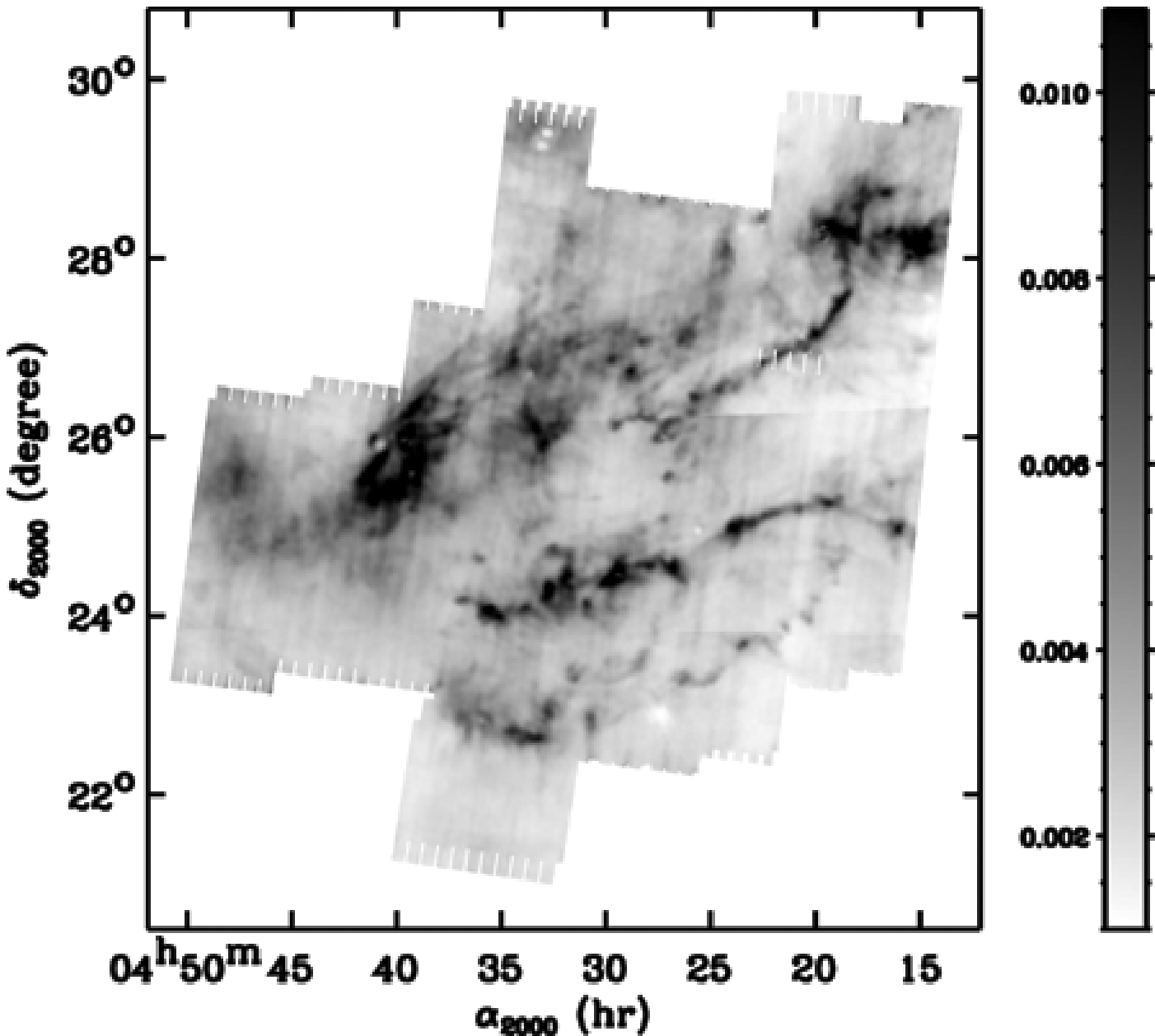}}
\caption{Comparison between (\textit{a}) the visual extinction map from \citet{Padoan2002} based on NIR observations and (\textit{b}) the dust opacity at 160 $\mu$m deduced from our dust temperature map.}
\label{fig:tau_and_av}
\end{figure*}

\begin{figure*}[!t]
\centering
\subfigure[]
	{\label{fig:emissivity}
	\includegraphics[angle=90,width=.3\linewidth]{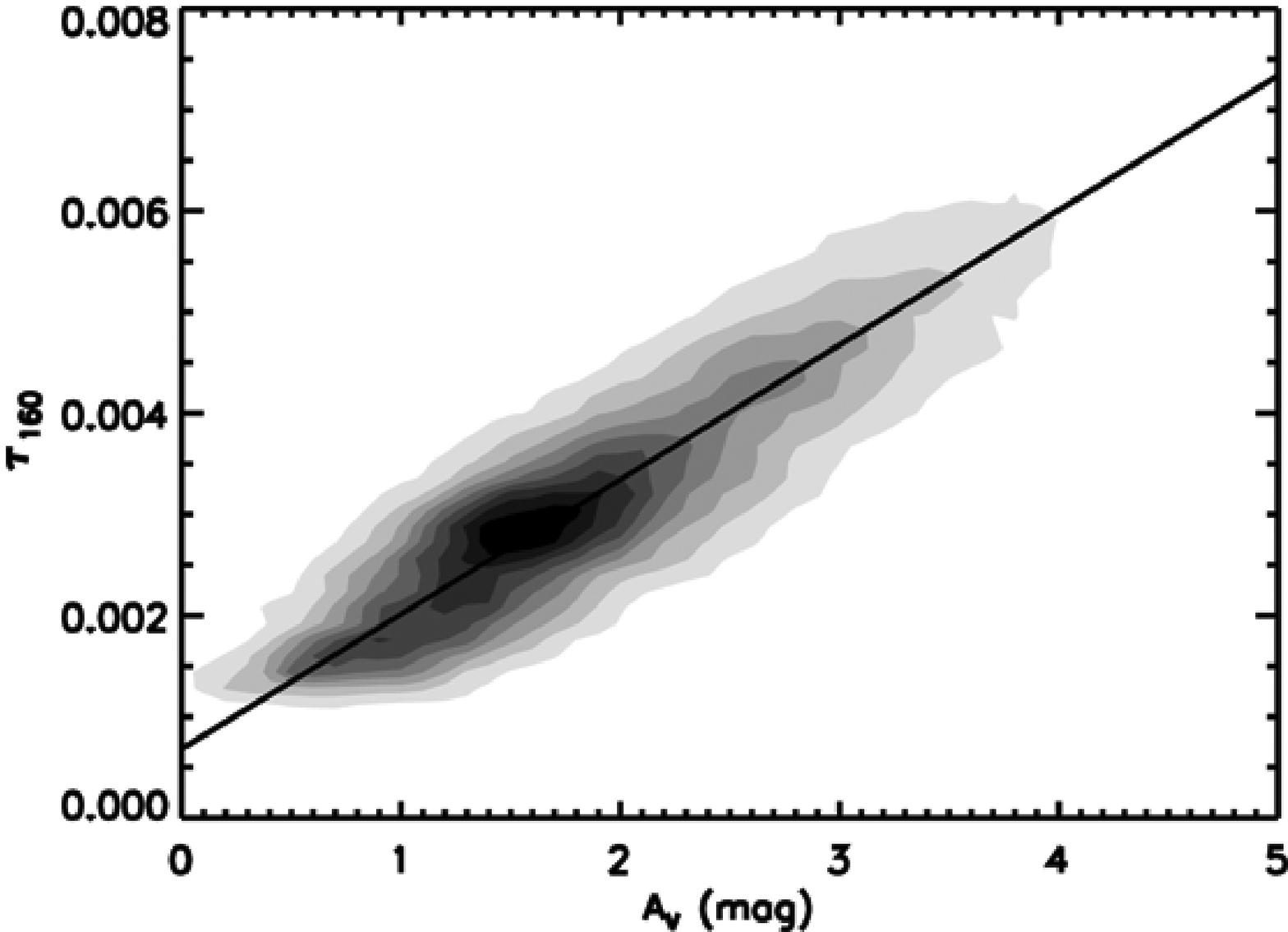}}
\subfigure[]
	{\label{fig:tau_av_tdust}
	\includegraphics[width=.3\linewidth]{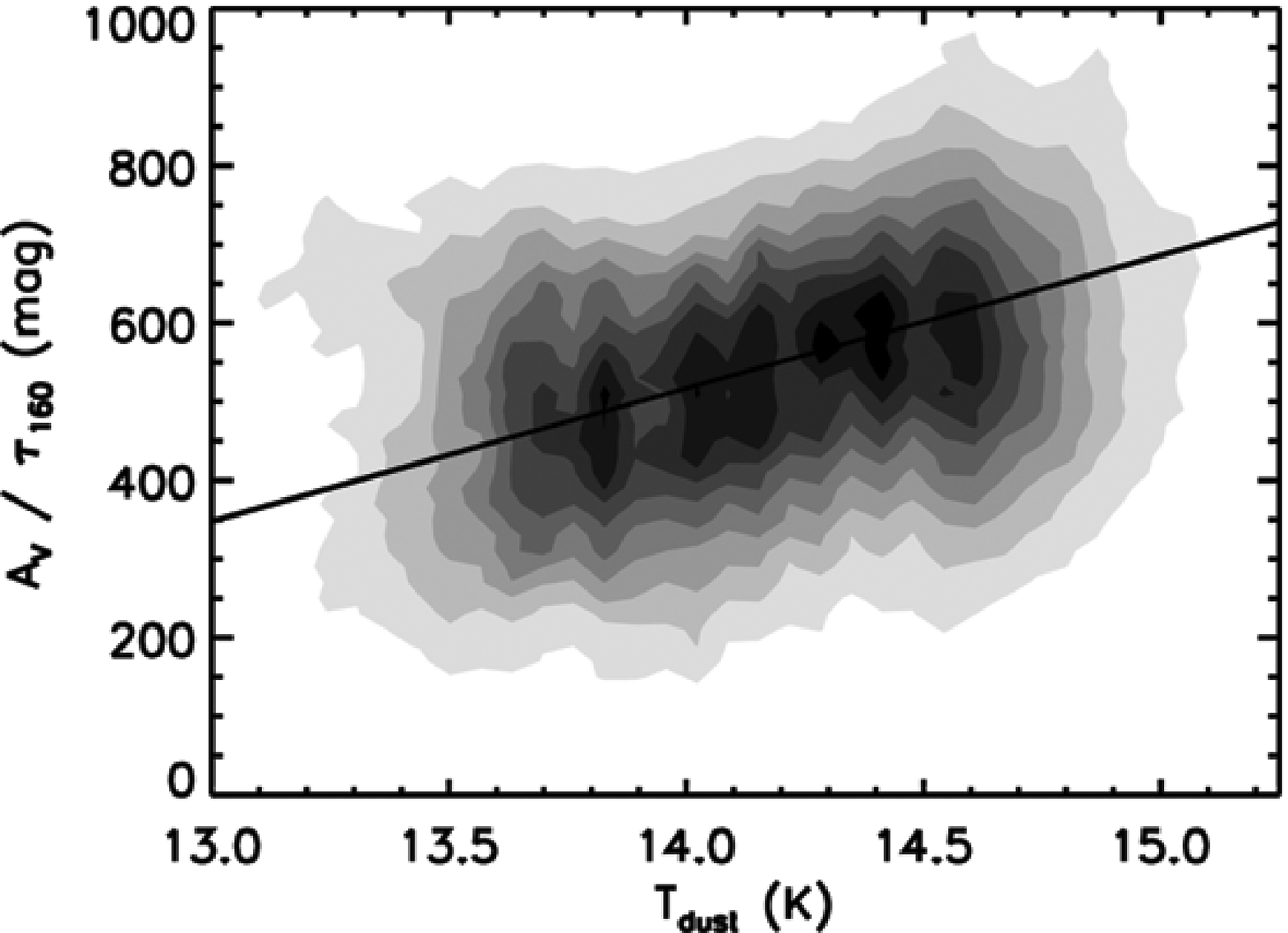}}
\subfigure[]
	{\label{fig:160av}
	\includegraphics[angle=90,width=.3\linewidth]{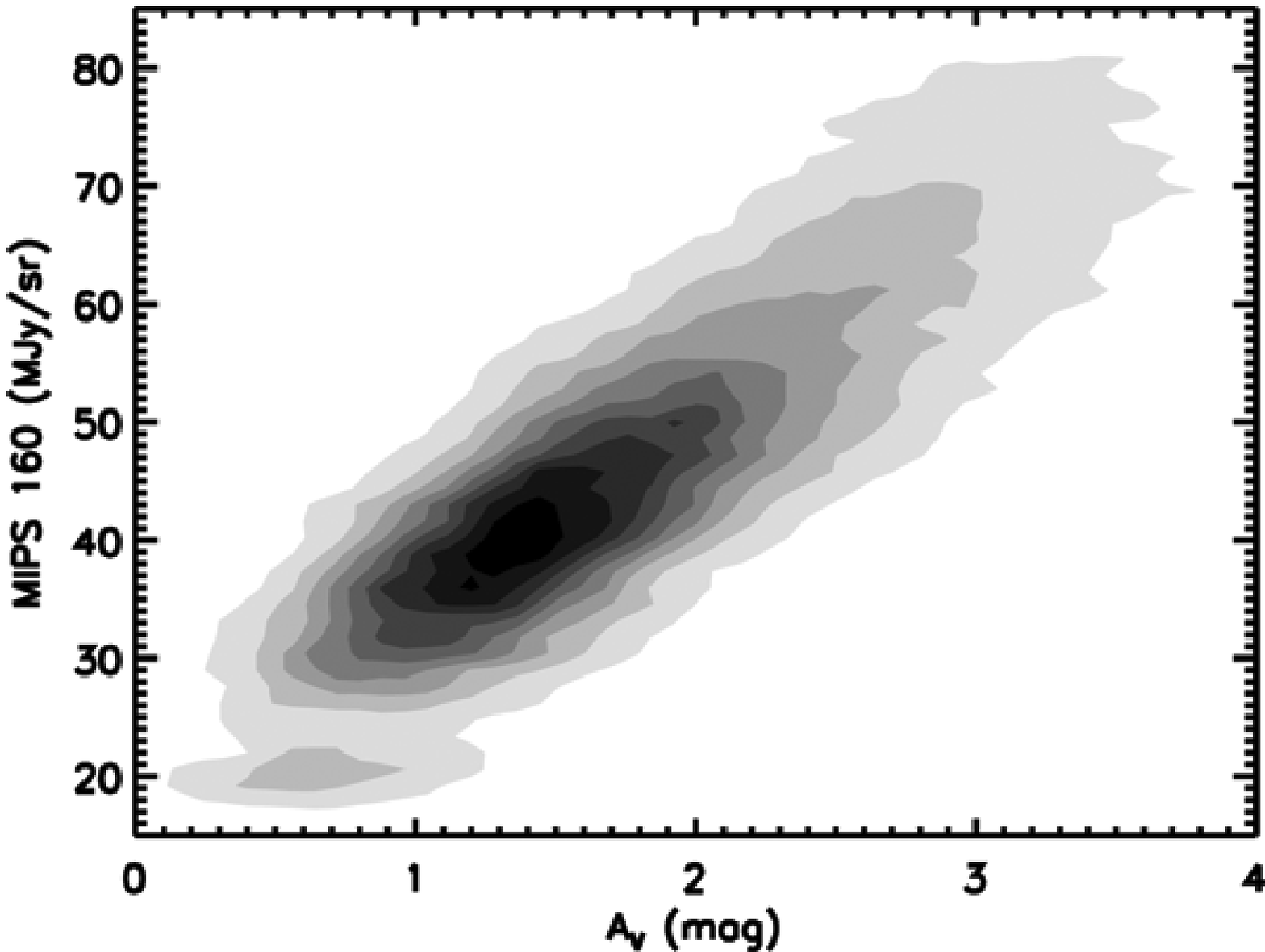}}
\caption{Correlation plot of (a) the FIR depth at 160 $\mu$m as a function of visual extinction, (b) the visual extinction to FIR depth ratio at 160 $\mu$m as a function of the BGs temperature and (c) the dust emission at 160 $\mu$m as a function of visual extinction.}
\label{fig:av_plots}
\end{figure*}

We use the dust temperature map to build FIR dust opacity maps. For every pixel of our dust temperature map we compute the Planck function and convolve it with the IRAS 100 $\mu$m or MIPS 160 $\mu$m transmission filter. We combine these values and the dust FIR emission at 100 or 160 $\mu$m to build maps of the dust opacity at 100 and 160 $\mu$m (see Fig.~\ref{fig:taumap}). \citet{Padoan2002} have computed the structure function scaling of a 2MASS extinction map of the TC complex using the methods proposed by \citet{Cambresy2002}. Not surprisingly, there is a strong correlation between their extinction map and our dust opacity maps (see Fig.~\ref{fig:tau_and_av}). From these correlations, we can compute the dust emissivity, assuming it follows a law in $\nu^2$. We find $\tau_{100}/A_v = (3.65\pm0.09)\times10^{-3}\ \rm{mag^{-1}}$ and $\tau_{160}/A_v = (1.34\pm0.04)\times10^{-3}\ \rm{mag^{-1}}$ (see Fig.~\ref{fig:emissivity}).

\citet{Boulanger1996} derived, from COBE and HI observations at high Galactic latitude, a diffuse dust emissivity that corresponds to $1.18\times10^{-3}\ \rm{mag^{-1}}$ and $0.46\times10^{-3}\ \rm{mag^{-1}}$ at 100 and 160 $\mu$m respectively using $A_V/N_H = 0.53 \times 10^{-21} \rm{cm^2}$ \citep{Savage1979}. There is a factor of about 3 between the cirrus measurements and those from the TC. However, if we assume the dust temperature is 17.5 K within the cirrus, as \citet{Boulanger1996} measured it on FIRAS observations, then their cirrus DIRBE colors from 100 to 240 $\mu$m are better adjusted with an emissivity law of $\nu^{1.73}$. If we use this law for the TC, we then find $\tau_{100}/A_v = (2.58\pm0.06)\times10^{-3}\ \rm{mag^{-1}}$ and $\tau_{160}/A_v = (1.08\pm0.04)\times10^{-3}\ \rm{mag^{-1}}$, a factor 2.2--2.3 above the cirrus value from \citet{Boulanger1996}. Variations of the hydrogen column density to visual extinction conversion $A_V/N_H$ could be the reason to explain the differences revealed by comparing our measurements, deduced from direct comparison to visual extinction, to those of \citet{Boulanger1996}, who have been using $\rm{HI}$ observations. However, FUSE observations have shown that this ratio does not change from diffuse to translucent clouds \citep{Rachford2002}.
\citet{DelBurgo2003} and \citet{Kiss2006} used ISOPHOT observation to measure the dust emissivity at 200 $\mu$m, $\tau_{200}/A_v = 0.28\times10^{-3}\ \rm{mag^{-1}}$ for lines of sight towards the TC, and $\tau_{200}/A_v = 0.2 - 1.0 \times10^{-3}\ \rm{mag^{-1}}$ for other galactic lines of sight. The difference between these estimates and our values might only arise from the wavelength dependance of the FIR dust emissivity. It might also be due to the fact that their numbers are based on observations most often covering a too small area to provide a background outside of the cloud.
An increase of the FIR emissivity by a factor of a few relative to the diffuse ISM has also been observed in the \textit{cold} component of the Taurus Molecular Cloud TMC-2 \citep{delBurgo2005}. They interpreted this increase as an indication that the transiently heated VSGs are not present within the \textit{cold} component but rather coagulate on bigger grains and that ice mantles form on dust.

Figure \ref{fig:tau_av_tdust} shows the variation of the visual extinction to FIR depth ratio at 160 $\mu$m as a function of the BGs temperature for one IRIS plate that covers the TC. The BGs emissivity significantly increases with decreasing dust temperature. The correlation slightly varies over the entire complex, from one IRIS plate to another. This may arise from the difference in the dust temperature range that is spanned within each plate. The IRIS plate that is used for the correlation plot of Fig.~\ref{fig:tau_av_tdust} is the one that exhibits the largest contribution of pixel below 14 K. This correlation between low BGs temperature and high FIR emissivity is in agreement with the results from \citet{Schnee2008} for the Perseus Molecular Cloud.

In order to confirm the good correlation between MIPS 160 $\mu$m data and the dense and cold structures in the cloud, we compare these FIR observations to the visual extinction map. On average, the resolution of the map from \citet{Padoan2002} is 3.4\arcmin\ but varies depending on the local number density of stars. We convolve the IR observations by a variable beam before the comparison is done. The correlation plot is shown on Figure \ref{fig:160av}. The most extinguished structures are those with the most FIR emission and are the cold filaments on the dust temperature map. The good correlation between dust emission at 160 $\mu$m, low BGs temperature and the extinction confirms that the dust FIR emission is a good tracer of dense regions within molecular clouds as it has previously been shown by \citet{Langer1989} and \citet{Abergel1994}.

\subsection{Dust modeling}
\label{lab:sed_mod}

We compare the average mid- to far-infrared SED of the TC to that of the Diffuse Galactic Medium from \citet{Flagey2007}. At these wavelengths, the Taurus dust emission has a steeper spectrum than in the diffuse galactic medium and this can be related to both different grain size distribution and temperature. We explore both of these interpretations using the dust model of \citet{Compiegne2008}, an updated version of the well-known model of \citet{Desert1990} to fit the average SED of the TC from 60 to 240 $\mu$m. The VSG and BG abundances and the interstellar radiation field (ISRF) intensity are free parameters in the model. Hereafter, we express the ISRF intensity in units of the Solar Neighborhood far-UV field: $1.6\times10^{-3}\ \rm{erg.s^{-1}.cm^{-2}}$ integrated from 912 to 2000 $\mbox{\AA}$ (Habing unit). The best fit model spectrum is shown as a solid line in Figure \ref{fig:seds}.

The corresponding ISRF is $\sim0.2$, which is significantly low and may not represent the real mean incident radiation field. Indeed, the model does not take into account the observed increase in the far-IR dust emissivity. Therefore, the lower dust temperature, as compared to that of the diffuse Galactic medium, is entirely accounted for by reducing the radiation field intensity. However, for a given grain temperature the power radiated by dust in the far-IR is proportional to the dust emissivity while the absorbed power scales with the dust absorption coefficient in the UV/optical and the radiation field intensity. The balance between absorption and emission implies that for a larger dust emissivity, the same grain equilibrium temperature is obtained for a proportionally larger absorbed power. Then, assuming that the UV/optical dust absorption coefficient does not change, the previously stated ISRF intensity of $\sim0.2$ would rather be in the range 0.5 to 0.6 if the model included the observed increase in far-IR dust emissivity.

For each dust component (PAHs, VSGs and BGs), the model emission scales with the quantity of emitting grains (i.e. the dust mass in each component) and the grain heating power (i.e. the radiation field intensity). For a given emission to be fitted, a low value of the radiation field translates in a higher dust mass in each dust component. As a consequence, the dust total abundance is overestimated, though the relative fractions of each dust component are well fitted. The abundance ratio of BGs and VSGs is about 1.1 times that of \citet{Desert1990} and 2.3 times that of \citet{Flagey2007} for the diffuse Galactic medium. However, this abundance ratio relies mostly on the 60 $\mu$m emission from VSGs. The increase of the FIR dust emissivity can be related to this difference in the VSGs to BGs relative abundance. Within a dark cloud of the TC, \citet{Stepnik2003} observed an increase of the sub-millimeter emissivity coupled with a decrease of the VSGs abundances, with factors relative to diffuse medium larger than ours, and interpreted them by introducing grain-grain coagulation into fluffy aggregrates as an important process inside the cloud. In the following section, we discuss evidence for spatial variations in PAHs and VSGs abundance.

\section{Local variations of MIR observations}
\label{lab:local}

\subsection{Selection of the sub region}
\label{lab:selec_3f}

\begin{figure*}[!t]
\centering
\subfigure[]
	{\label{fig:3f_map8}
	\includegraphics[width=.425\linewidth]{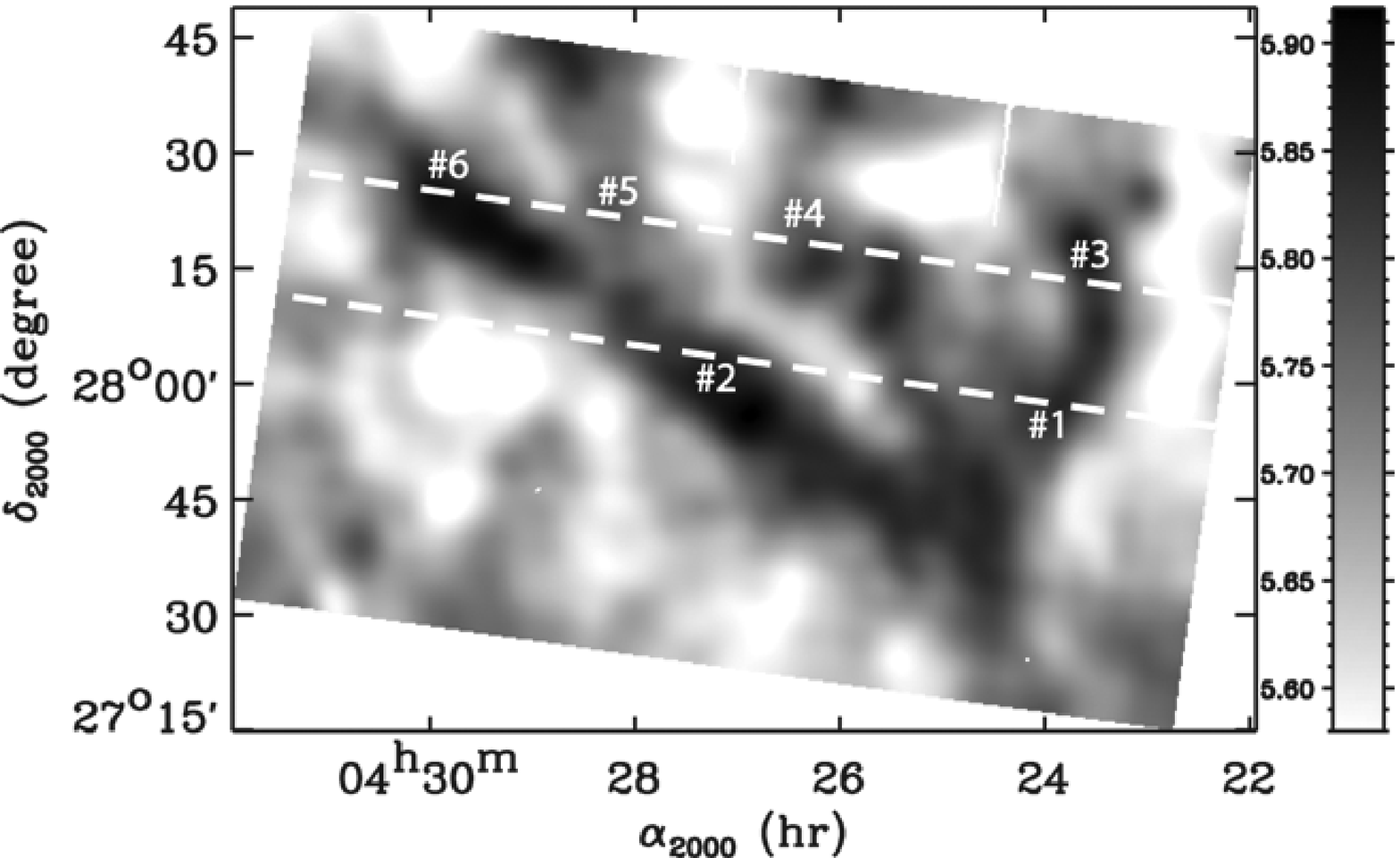}}
\subfigure[]
	{\label{}
	\includegraphics[angle=90,width=.425\linewidth]{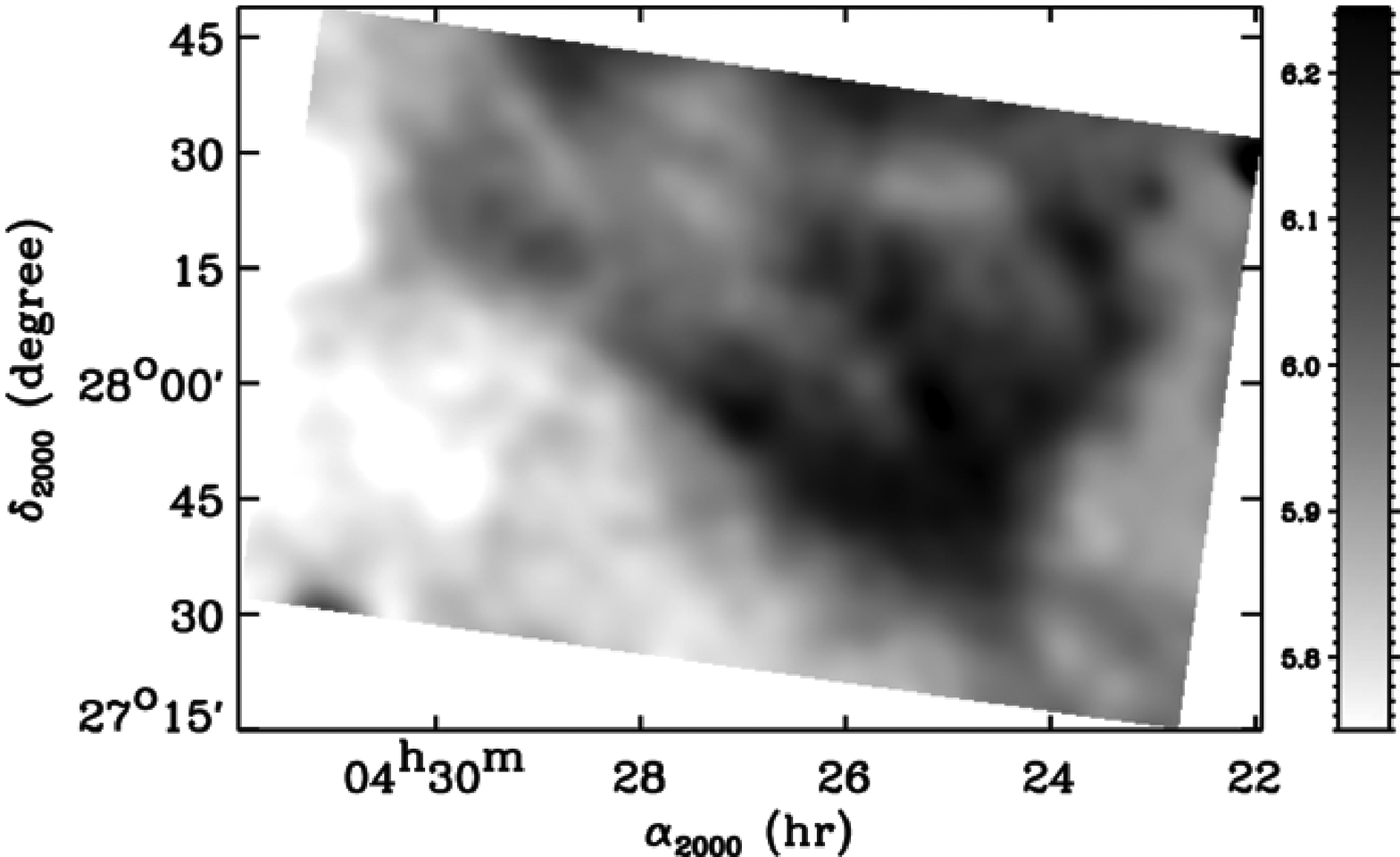}}
\subfigure[]
	{\label{}
	\includegraphics[angle=90,width=.425\linewidth]{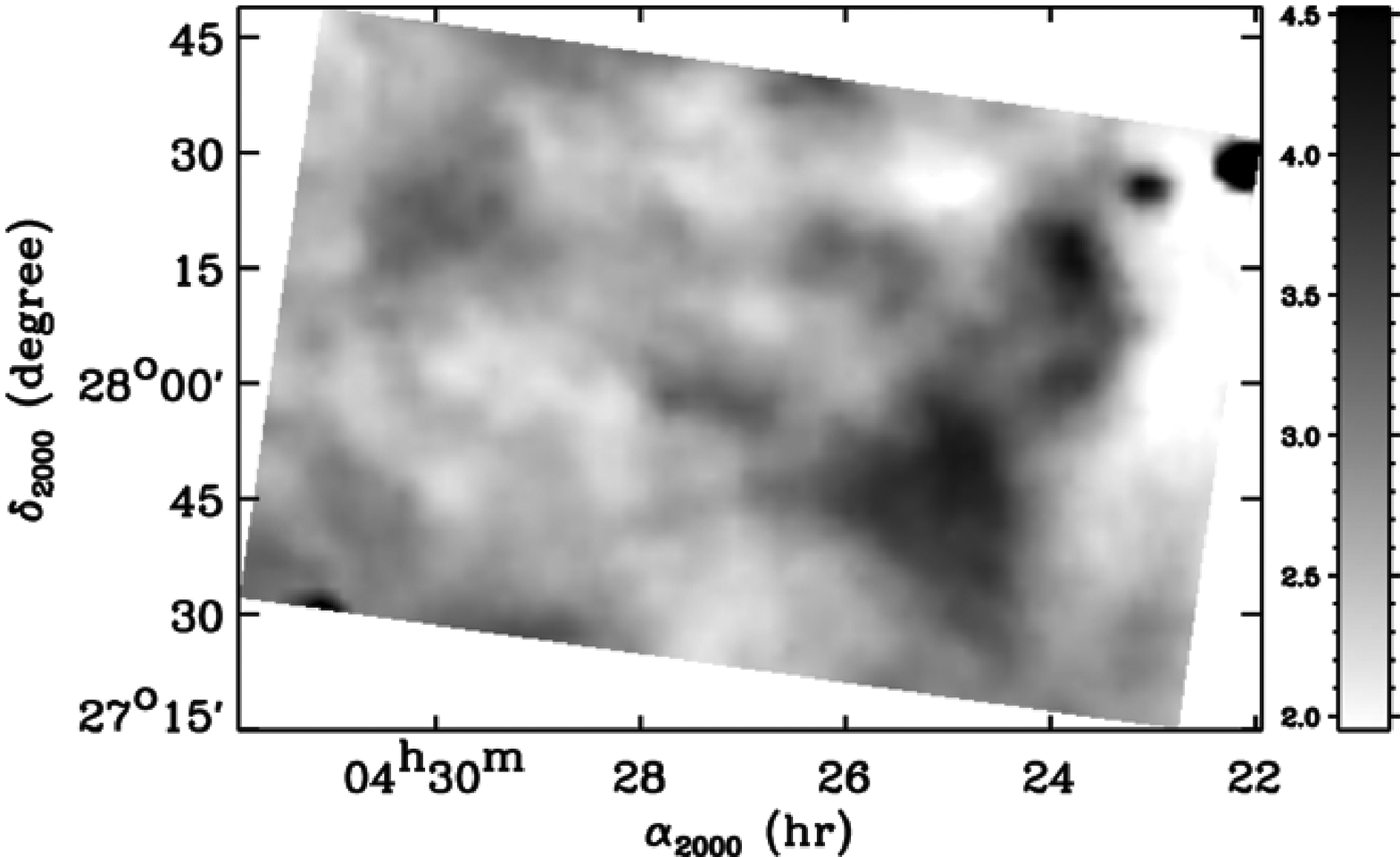}}
\subfigure[]
	{\label{}
	\includegraphics[angle=90,width=.425\linewidth]{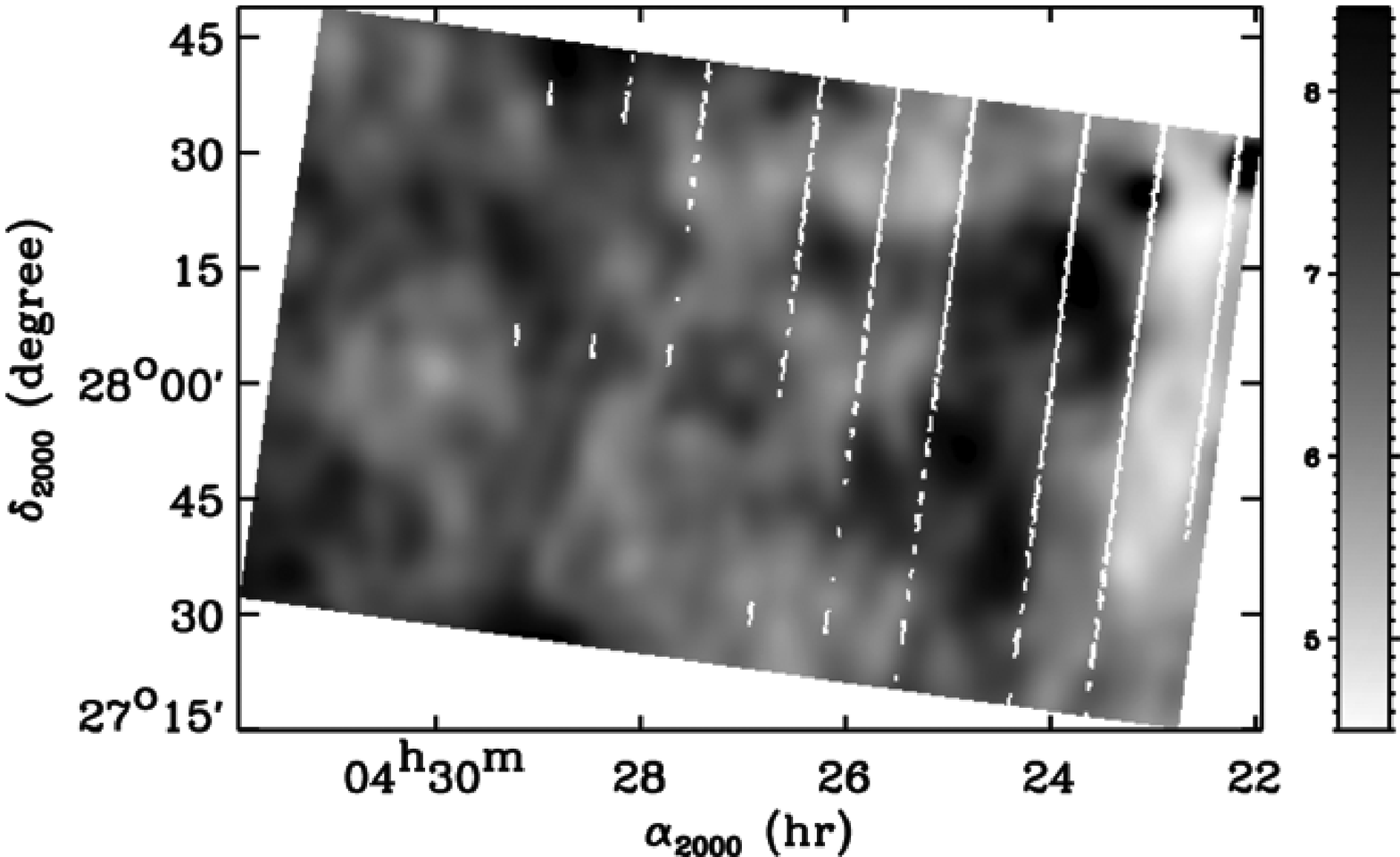}}
\subfigure[]
	{\label{}
	\includegraphics[angle=90,width=.425\linewidth]{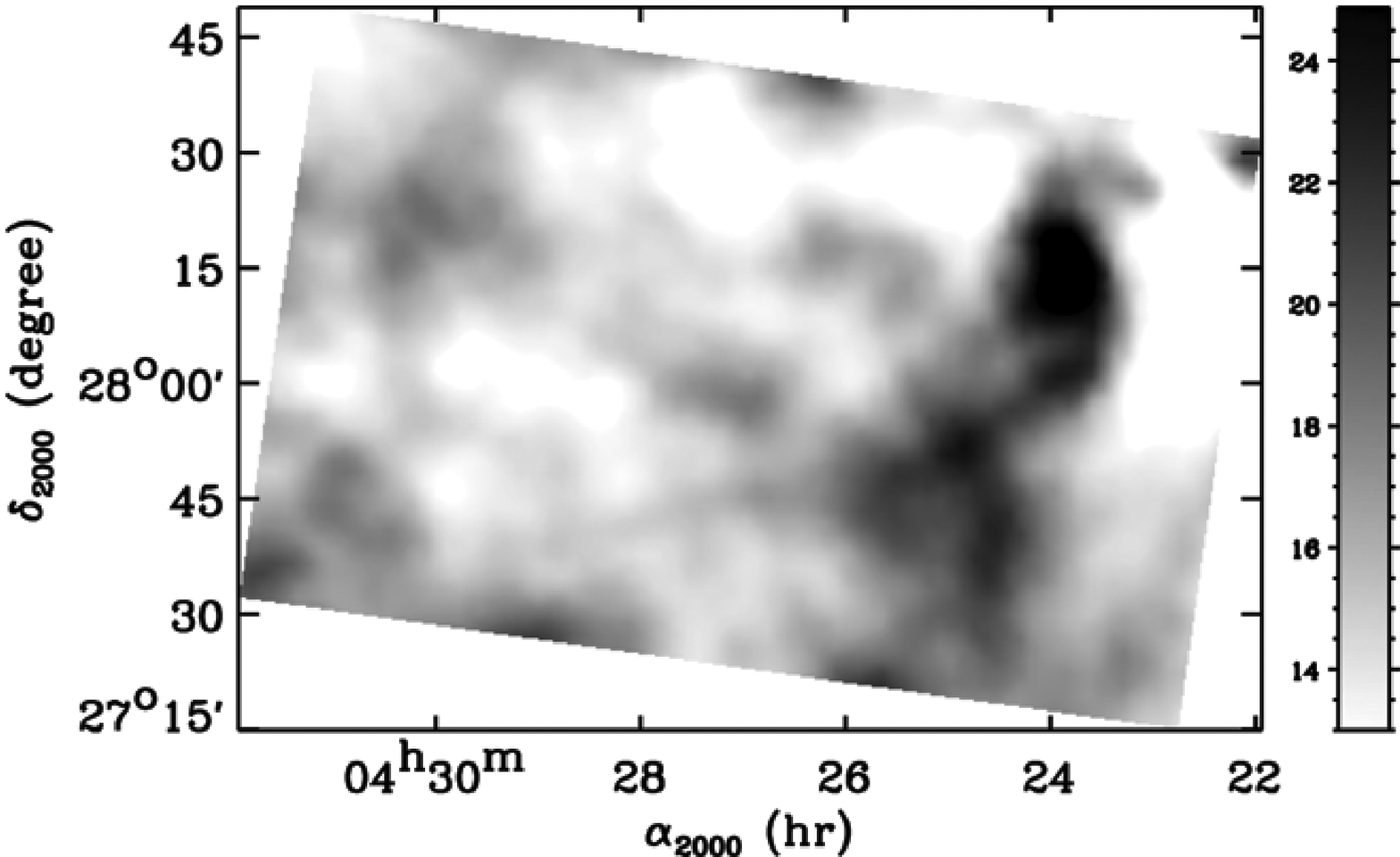}}
\subfigure[]
	{\label{}
	\includegraphics[angle=90,width=.425\linewidth]{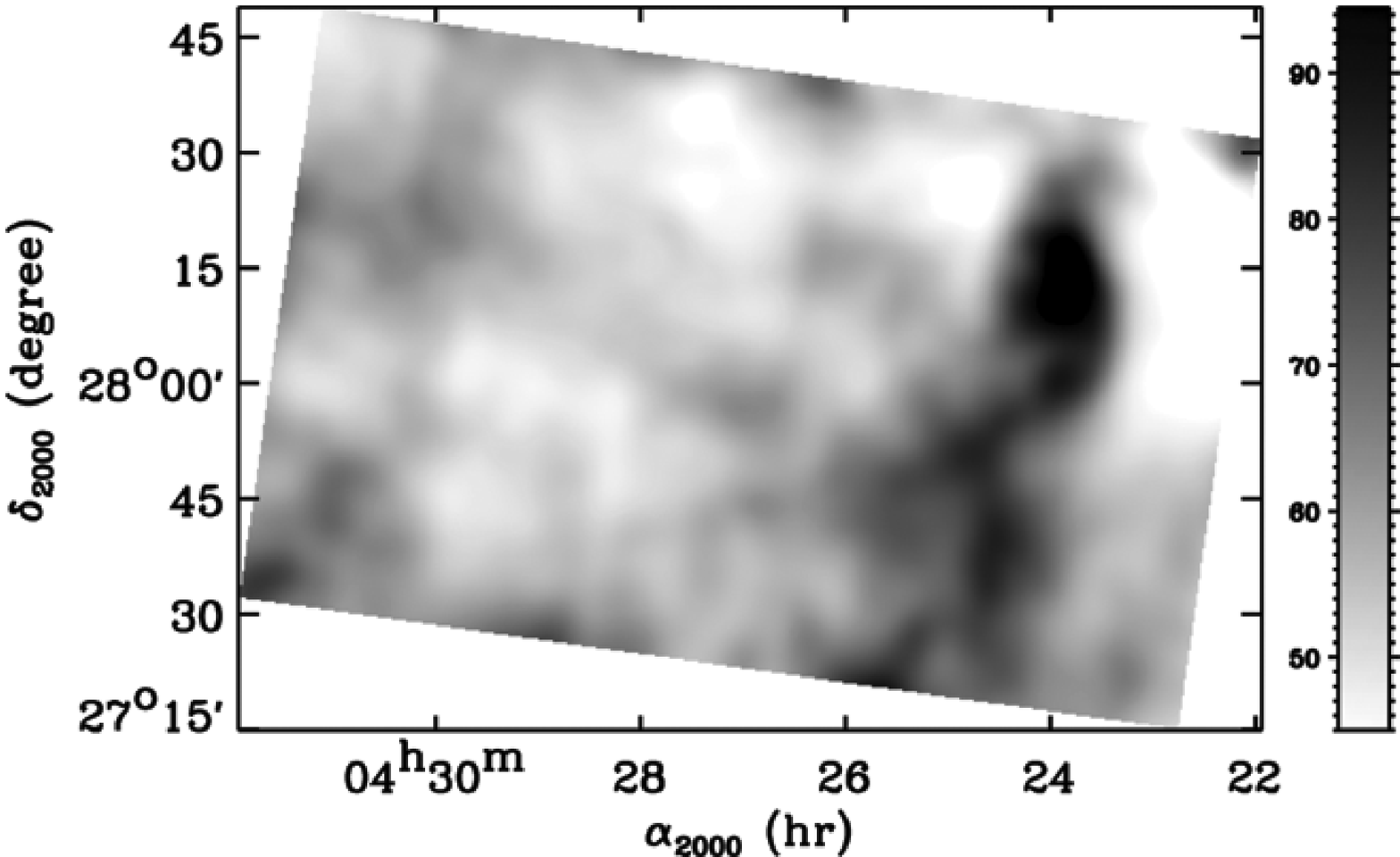}}
\caption{Observations from 8 to 160 $\mu$m of the "3 fingers" sub-region at the IRAS 100 $\mu$m resolution of 4.3\arcmin: (\textit{a}) IRAC 8 $\mu$m, (\textit{b}) MIPS 24 $\mu$m, (\textit{c}) IRIS 60 $\mu$m, (\textit{d}) MIPS 70 $\mu$m, (\textit{e}) IRIS 100 $\mu$m, and (\textit{f}) MIPS 160 $\mu$m. The dashed lines correspond to the two cuts made across the filamentary structures, which are shown on Figure \ref{fig:3f_cuts}. The units are in MJy/sr.}
\label{fig:3f_maps}
\end{figure*}

\begin{figure*}[!t]
\centering
\subfigure[]
	{\label{fig:3f_cut_a}
	\includegraphics[angle=270,width=.475\linewidth,angle=180]{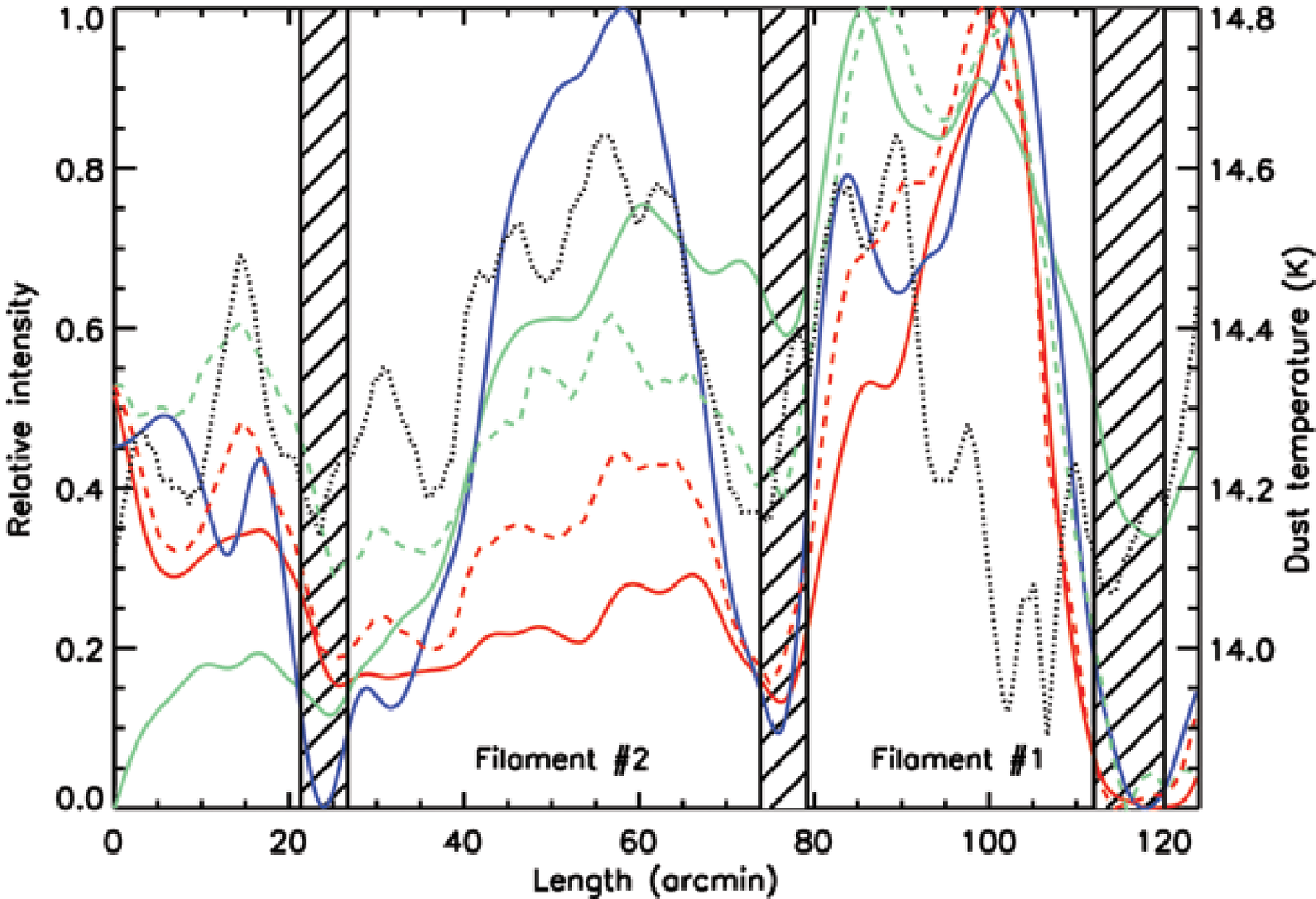}}
\hspace{.0125\linewidth}
\subfigure[]
	{\label{fig:3f_cut_b}
	\includegraphics[angle=270,width=.475\linewidth,angle=180]{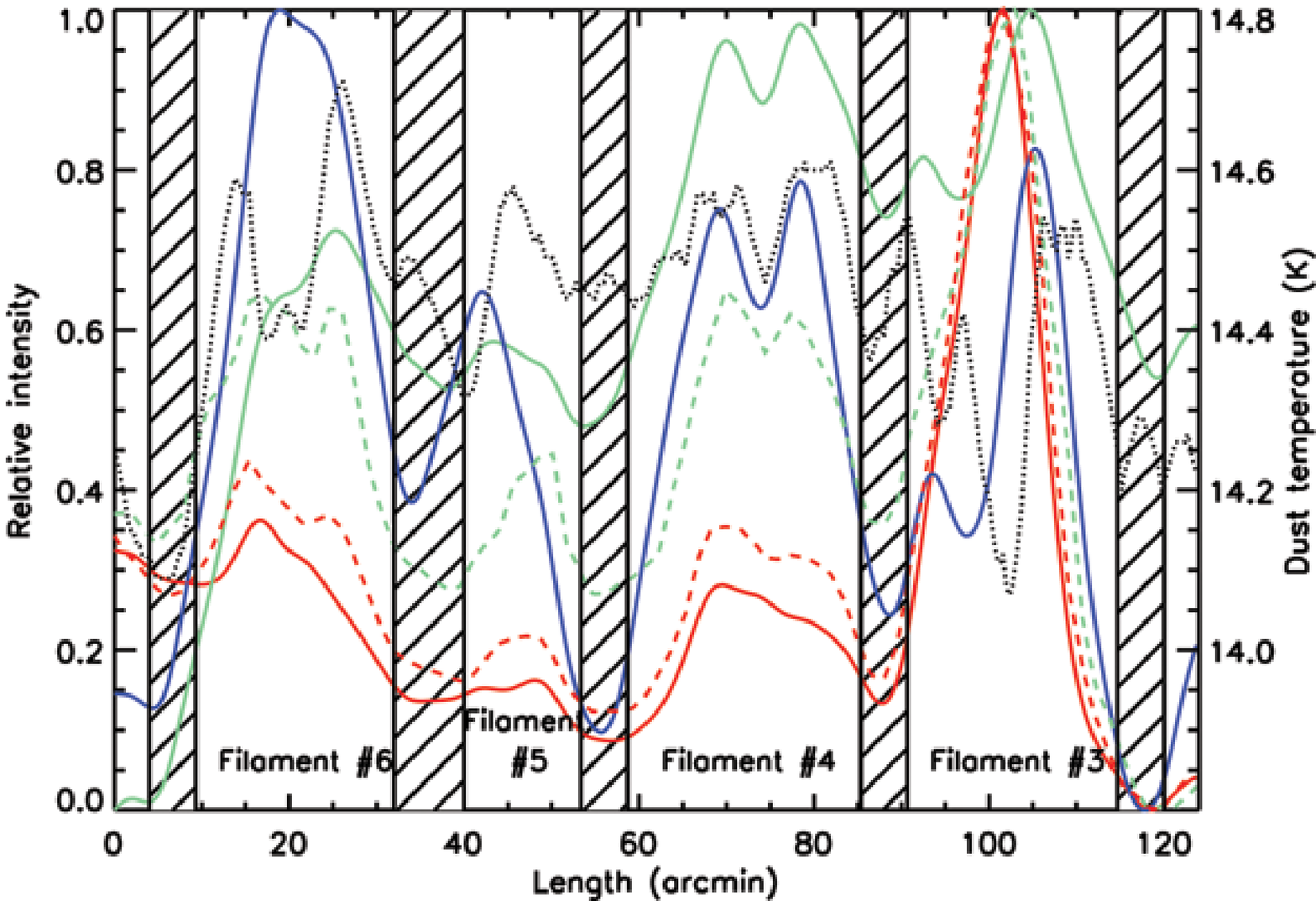}}
\caption{Relative MIR and FIR intensities along the two cuts as shown on Figure \ref{fig:3f_maps}. (\textit{a}) is the cut \#1 and (\textit{b}) is the cut \#2. Continuous lines are \textit{Spitzer} values: (\textit{red}) MIPS 160 $\mu$m, (\textit{green}) MIPS 24 $\mu$m and (\textit{blue}) IRAC 8 $\mu$m. Dashed lines are IRIS values: (\textit{red}) IRIS 100 $\mu$m and (\textit{green}) IRIS 60 $\mu$m. Dotted lines are the temperature as deduced from the BG temperature map (Fig.~\ref{fig:dusttempmap}). The hatched areas represent the borders of each filament where the local minima have been measured to estimate the background.}
\label{fig:3f_cuts}
\end{figure*}

Zodiacal light is the dominant component of the diffuse emission at 8 and 24 $\mu$m and may contribute significantly to the observed spatial variations. We select a region, 2.6 square degrees in extent, away from the brightest zodiacal bands, that exhibits filamentary structure at both 8 and 24 $\mu$m (see Fig.~\ref{fig:3f_maps}). At the shortest wavelengths (8 and 24 $\mu$m), three filaments are clearly visible, forming a "three fingers hen-foot" shape. At longer wavelengths (60 $\mu$m and above), one of these filaments is significantly brighter than the two others, to the extent that it remains the one fully visible at 100 and 160 $\mu$m. This "3 fingers" structure is also visible at the same position within IRIS 12 and 25 $\mu$m images (see Fig.~\ref{fig:data_12_8} and \ref{fig:data_25_24}). We thus assume it is part of the TC and not of the foreground zodiacal emission.

We perform two cuts through the filaments. The positions of these cuts are shown in Figure \ref{fig:3f_maps} and their orientation is from high to low right ascension. Each map is at the resolution of IRAS 100 $\mu$m (4.3\arcmin) so we take a one pixel wide straight line for each cut. The variation of MIR and FIR emission along these cuts is shown on Figure \ref{fig:3f_cuts}. On each cut, we identify several filaments thanks to the common presence of peak at every wavelength, from 8 to 160 $\mu$m. The relative position and amplitude of these peaks significantly change from one wavelength to another, from one peak to another. From east to west, we label filaments \#1 and \#2 on the first cut, filaments \#3, \#4, \#5 and \#6 on the second cut (see Fig.~\ref{fig:3f_map8} to identify them on the IRAC 8 $\mu$m image). If we look back at the observations of this region, we see that filament \#2 and filament \#6 appears to be the same "finger" or spatial feature, while the "finger" or structure that contains filament \#1 splits into two sub-features that correspond to both filament \#3 and filament \#4. Filament \#5 seems to be part of a structure connected to filament \#6, judging by the IRAC 8 $\mu$m observations. Projection effects may significantly affect such associations.

\subsection{Average colors of the sub-region}

We measure the MIR to FIR colors of this field as we did previously on the entire cloud (see Section \ref{lab:sed}). At longer wavelengths (at 60 $\mu$m and above), the correlations are still good, except when MIPS 70 $\mu$m is involved because of the significant noise of the data. At shorter wavelengths (IRAC 8 vs MIPS 24 and MIPS 24 vs IRIS 60), the dispersion is still significantly higher preventing a clean correlation and measurement of colors for those wavelengths. The measured FIR colors within this selected region are shown in Table \ref{tab:3f_colors} as well as those for the whole TC. The FIR colors of the "3 fingers" sub-region are globally in agreement with those from the entire TC.

\subsection{Small spatial scale variations in dust SEDs}

\begin{figure}[!t]
	\centering
	\includegraphics[angle=90.,width=0.75\linewidth]{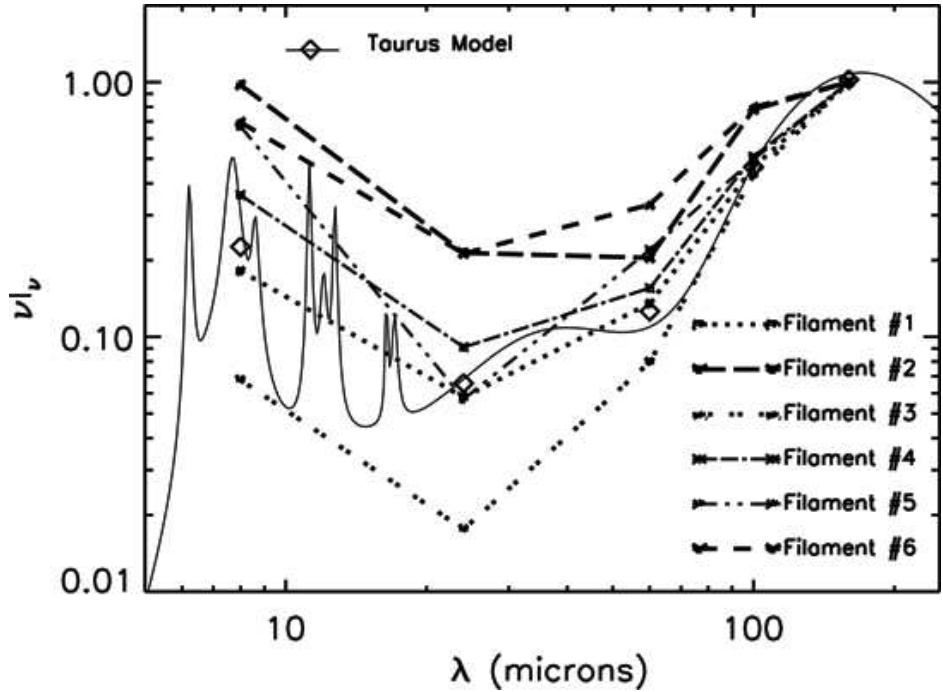}
	\caption{Spectral energy distribution of the filaments within the "3 fingers" sub-region of the TC, from 8 to 160 $\mu$m, normalized at 100 $\mu$m. Reference model spectra for the Galactic diffuse medium and the TC are also plotted.}
	\label{fig:3f_sed}
\end{figure}

In order to constrain the variations of the dust properties on small spatial scales, we measure the colors of each filament independently. We first determine the borders of each of these structures on the cuts by looking for the minima values within the hatched areas (see Fig.~\ref{fig:3f_cuts}). For a given filament, the minima position may change from one wavelength to another by at most 2\arcmin. We then interpolate a straight line between the east and west borders to estimate the background that we remove from the filament intensity. We finally integrate the background  subtracted intensity between the two borders of the filament. The final colors of the filaments, relative to MIPS 160 $\mu$m, are shown in Table \ref{tab:3f_cuts_colors} and compared to the TC reference model in Fig.~\ref{fig:3f_sed}. For each filament, we give the background-subtracted brightness within the MIPS 160 $\mu$m channel as well as the corresponding visual extinction. The uncertainties on our measurements are dominated by the photometric uncertainties of the observations: 10\% on IRAC8 and MIPS24 and 15\% on MIPS160, IRIS60 and IRIS100 \citep{Reach2005, Engelbracht2007, Gordon2007, Stansberry2007}. We give these absolute photometric errors but they are not relevant for the relative analysis of small-scale variations. We rather draw the reader's attention on the relative errors, which are much smaller. For instance, the standard deviations at 8 and 24 $\mu$m, which include effects from structure and optical depth, are about 0.08 and 0.13 MJy/sr respectively, a factor 70 and 50 below the mean brightness at these wavelengths.

The filaments colors clearly show that there is a difference between filaments \#1, \#3 and \#4 as a group and filaments \#2 and \#6 as another. The first three present IRIS100/MIPS160 and IRIS60/MIPS160 colors similar to those of the global TC (see Table \ref{tab:3f_colors}). The last two present IRIS100/MIPS160 and IRIS60/MIPS160 a factor from 1.5 to 3 above those average colors. At shorter wavelengths, the difference between these two sets of filaments increases as the IRAC8/MIPS160 and MIPS24/MIPS160 colors change by a factor from 2 up to 14. Filaments \#2 and \#6 present the highest colors at these wavelengths too. The changes we observe in the IR colors of these filaments occur within low opacity structures. The visual extinctions are at most a few magnitudes (see Table \ref{tab:3f_colors}).

The difference at longer wavelengths can be related to different BGs temperatures, filaments \#2 and \#6 being "hotter" than the others, as shown in Figure \ref{fig:3f_cuts}. There is a difference of about 0.5 K between the structure that corresponds to filaments \#2 and \#6 and the one that includes filaments \#1 and \#3. This is in agreement with what we expect from our previous analysis of the FIR brightness evolution as a function of dust temperature on the whole molecular cloud. Filament \#5 appears to be in between the two groups of filaments but, as it is the faintest of our sample and as it presents a significant shift regarding the peak position, especially at 8 $\mu$m, we think it is not a conspicuous result and will not discuss it any further.

\section{Discussion}
\label{lab:disc}
\label{lab:colvar}

\begin{figure*}[!t]
\centering
\subfigure[]
	{\label{fig:60mdl100}
	\includegraphics[angle=90,width=.425\linewidth]{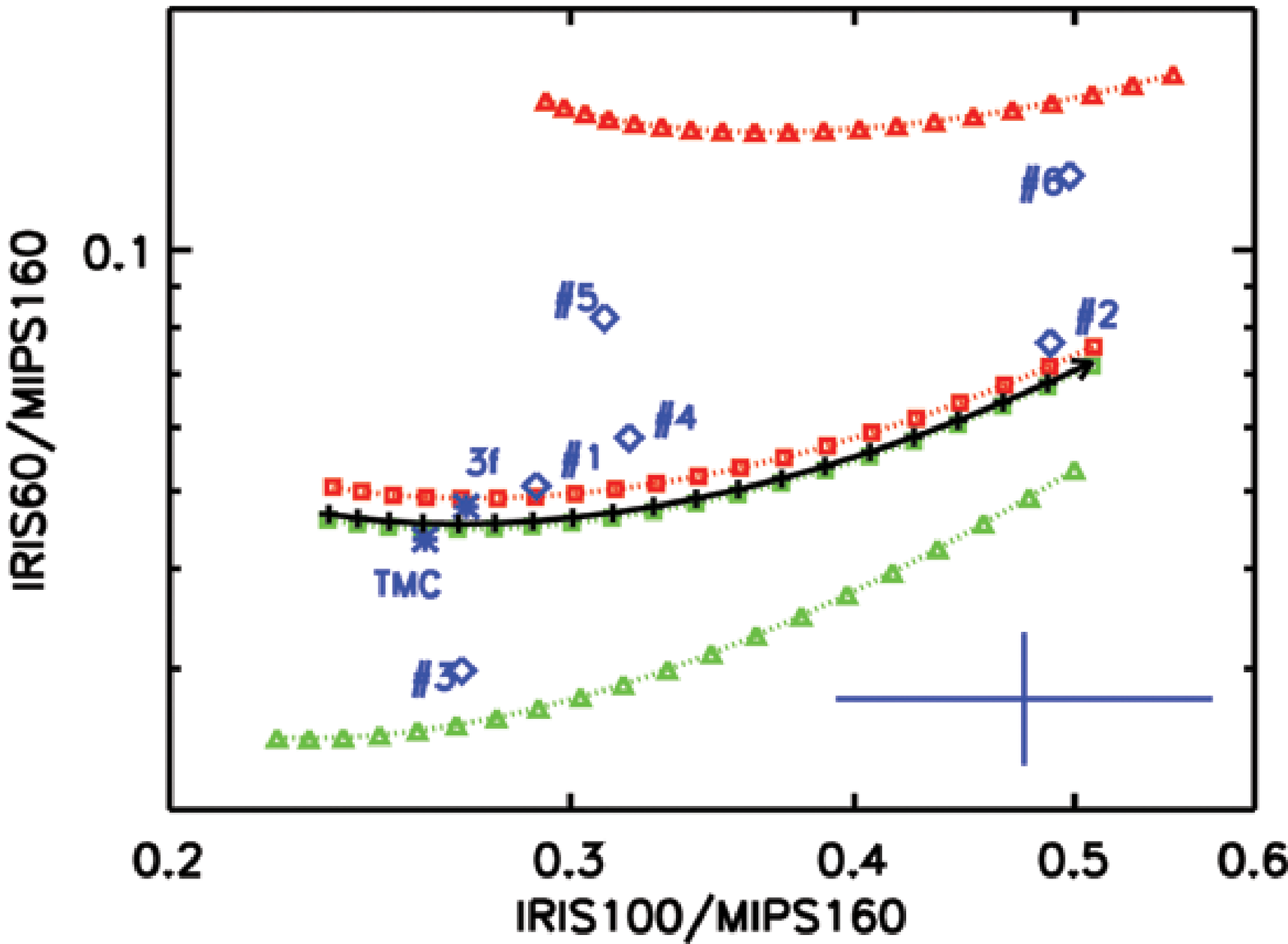}}
\hspace{.0125\linewidth}
\subfigure[]
	{\label{fig:24mdl100}
	\includegraphics[angle=90,width=.425\linewidth]{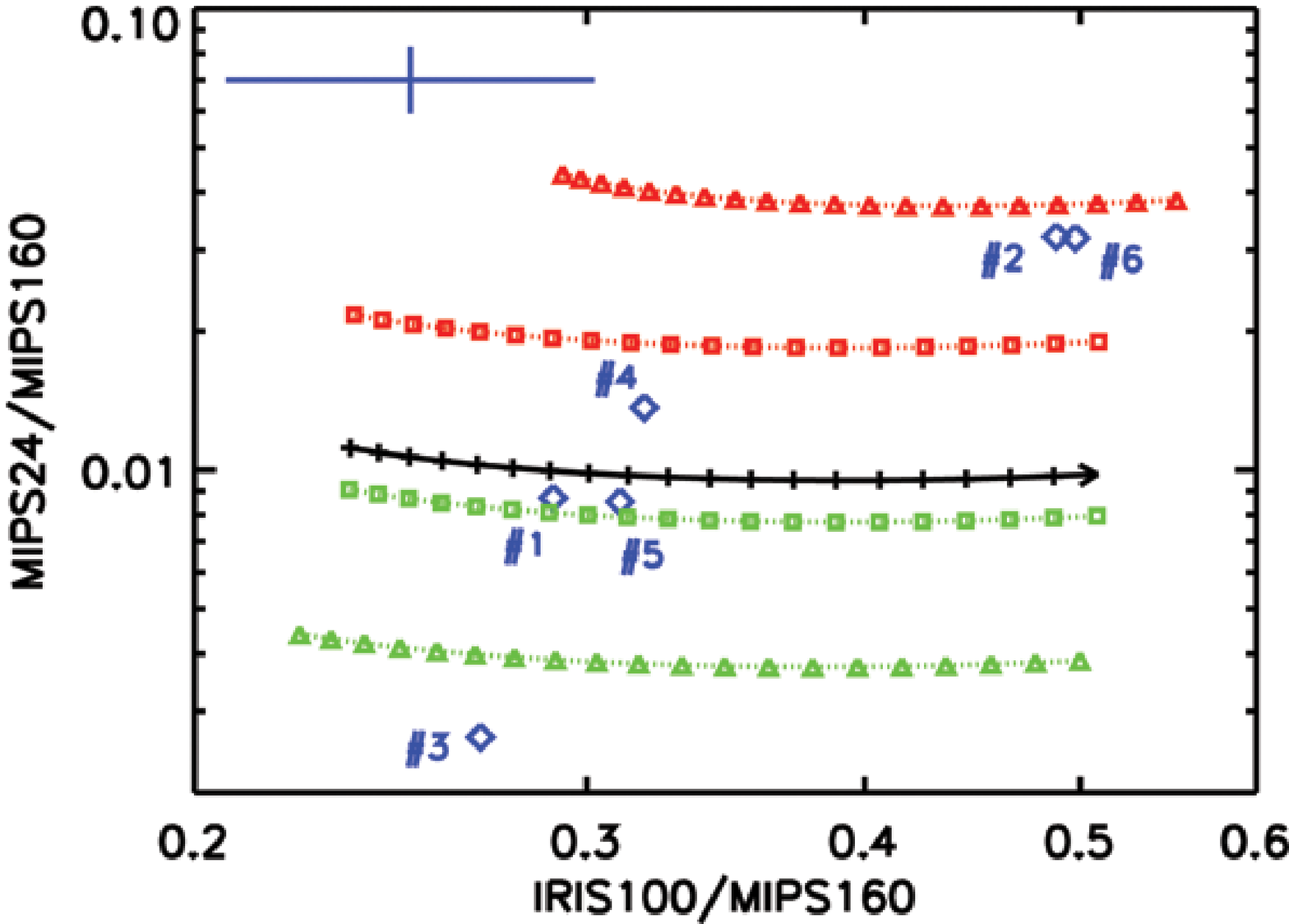}}
\subfigure[]
	{\label{fig:8mdl100}
	\includegraphics[angle=90,width=.425\linewidth]{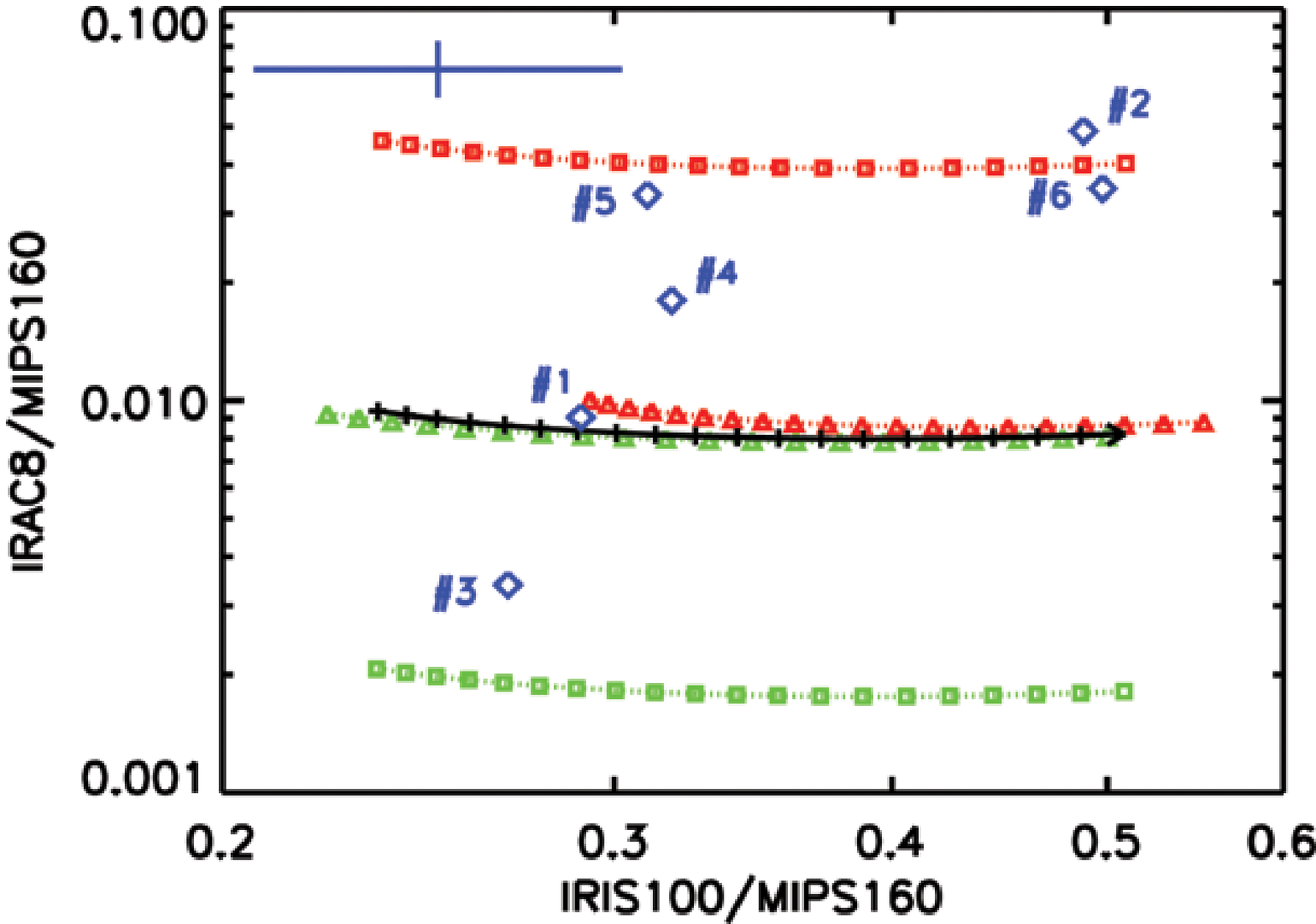}}
\caption{Color-color diagrams from the model of \citet{Compiegne2008} as a function of the ISRF intensity, VSGs and PAHs relative abundances. (\textit{a}) is IRIS60/MIPS160 vs IRIS100/MIPS160, (\textit{b}) is MIPS24/MIPS160 vs IRIS100/MIPS160, (\textit{c}) is IRAC8/MIPS160 vs IRIS100/MIPS160. The arrowed lines show the evolution of the colors as a function of increasing ISRF (from $\chi=0.3$ to $\chi=2.8$) for TC average dust abundances. The dotted lines show the same evolution for a factor 5 larger (\textit{red}) or lower (\textit{green}) relative abundance of the PAHs (\textit{squares}) or VSGs (\textit{triangles}). Blue diamonds are the colors measured on the filaments. Blue stars on (\textit{a}) are the average colors measured on the entire cloud and on the "three fingers" sub-area. The large blue crosses give the error-bars.}
\label{fig:model}
\end{figure*}

The color variations observed across the TC may result from either a change in dust relative abundances or in the excitation efficiency. We explore these two possibilities by comparing the measured colors of the filaments to those of the dust model. Our aim is to understand whether the variations of IR colors within the filaments can be related to variations of only interstellar radiation field (ISRF) intensity $\chi$ or if dust size distribution variations are required. In the following section, we use the average SED of the TC as a reference, both in terms of the ISRF intensity and dust size distribution, which is described by the relative abundances of PAHs, VSGs and BGs. As diagnostics, we use color-color diagrams that combine IRAC8/MIPS160, MIPS24/MIPS160, IRIS60/MIPS160 and IRIS100/MIPS160 ratios (see Fig.~\ref{fig:model}). Colors predicted by the model as well as colors measured within the TC are plotted in these diagrams.

\subsection{Variations of the ISRF intensity}

We first attempt to interpret the observed color variations of the filaments by varying $\chi$ from 0.3 and 2.8 times the reference ISRF. The dust size distribution is that of the average SED of the TC for the VSG and BG abundances (hereafter $Y_{VSG,0}$ and $Y_{BG,0}$). The PAH abundance (hereafter $Y_{PAH,0}$) is arbitrarily fixed at $Y_{PAH,0}/Y_{BG,0} = 4.0\times10^{-2}$ so that the IRAC8/MIPS160 ratio predicted by the model is within the range of observed IRAC8/MIPS160 ratios. As a comparison, the original PAH abundance in the model of \citet{Desert1990} is $Y_{PAH,0}/Y_{BG,0} = 6.7\times10^{-2}$. We show the model IR colors as a function of $\chi$ with black arrowed line in the color-color diagrams. The IRIS100/MIPS160 color is the most sensitive to the variations of $\chi$. It varies by a factor $\sim 2.5$, while IRIS60/MIPS160 varies by a factor $\sim 1.4$ (see Fig.~\ref{fig:60mdl100}). MIPS24/MIPS160 and IRAC8/MIPS160 colors are almost independent of $\chi$ (see Fig.~\ref{fig:24mdl100} and \ref{fig:8mdl100}).

The measured colors of the entire TC and the "3 fingers" sub-region are in agreement with those of the model. However, filament \#1 is the only one whose colors match in every diagram, within uncertainties, the modeled colors for the TC average dust abundances. A few other filaments colors (e.g. IRIS60/IRIS100 and IRIS100/MIPS160 ratios of filaments \#4 and \#2 on Fig.~\ref{fig:60mdl100}) are in agreement with the model and only require different values of $\chi$. The $\chi$ variations may be related to the extinction variations within the sub-region. However, the observed dispersion of the other filaments colors cannot be interpreted as due to variations of the ISRF intensity only.

The variations of $\chi$ may be responsible to modifications of the PAHs electric charge that can be traced by IRAC8 color \citep{Flagey2006}. Within the range of $\chi$ used here, we have estimated the variations of the IRAC8/MIPS160 color due to changes of the PAHs charge to be 4\% at most. Therefore, they are small relative to those observed in Fig.~\ref{fig:8mdl100} and we rule out this effect as a cause of the IRAC8/MIPS160 variations.

\subsection{Variations of the dust size distribution}

We change the dust size distribution by multiplying or dividing $Y_{PAH,0}$ or $Y_{VSG,0}$ by a factor 5. In order to disentangle between the effects of PAHs abundance variations and VSGs abundance variations, we modify only one abundance at a time ($Y_{PAH}$ or $Y_{VSG}$) while the two others ($Y_{VSG}$ and $Y_{BG}$ or $Y_{PAH}$ and $Y_{BG}$ ) are fixed. We thus create four different dust size distributions for which the abundance ratios relative to the reference model ($Y_{PAH}/Y_{PAH,0}:Y_{VSG}/Y_{VSG,0}:Y_{BG}/Y_{BG,0}$) are (5:1:1), (1:5:1), (1/5:1:1) and (1:1/5:1). We compute the IR colors for each size distribution as a function of $\chi$ and plot them in Figure \ref{fig:model}. All but one of the observed IR colors are within the range of values that the modeled colors span with the modified abundances.

Not surprisingly, the IRIS100/MIPS160 color is almost independent of the dust size distribution variations. It is unaffected by a change of the PAHs abundance and is only slightly modified by a change of the VSGs abundance. The variations of the dust size distribution are more significantly traced by the other colors. The IRIS60/MIPS160 color appears to directly trace the variations of the VSGs abundance relative to BGs and is almost independent of the PAHs abundance variations (see Fig.~\ref{fig:60mdl100}). However, the precise contribution of the VSGs to the IRIS 60 $\mu$m is still debatable. Besides, there may be different types of BGs with different equilibrium temperatures (see Section \ref{lab:dusttemp}). Some of them could have a slightly higher equilibrium temperature than the single temperature that we assume in the model and thus could make a significant fraction of the IRIS 60 $\mu$m emission. This statement does not apply to the emission in the MIPS 24 $\mu$m channel, which is only due to stochastically heated grains. In the TC, the contribution of the BGs to MIPS 24 $\mu$m emission is negligible. The MIPS24/MIPS160 color is mainly dependent on the VSGs abundance variations even though it is not completely independent of the PAHs abundance variations (see Fig.~\ref{fig:24mdl100}). The last diagram shows that the IRAC8/MIPS160 color is almost entirely dependent on the PAHs abundance variations (see Fig.~\ref{fig:8mdl100}). We use these two last colors as tracers of the PAHs and VSGs abundance variations within the TC. An increase of the VSGs abundance is required for filaments \#6, \#2 and to a lesser extent \#4; a decrease of the VSGs abundance is required for filament \#3; no change of the VSGs abundance is required for filament \#1 and \#5. An increase of the PAHs abundance is required for filaments \#2, \#6, \#5 and to a lesser extent \#4; a decrease of the PAHs abundance is required for filament \#3; no change of the PAHs abundance is required for filament \#1.

Thus, the colors of filaments \#2, \#3, \#4 and \#6 require an increase/decrease of both the PAHs and VSGs abundances while colors of filament \#1 are in agreement with the reference size distribution. It is clear that one has to invoke PAHs and VSGs abundance variations by a factor of a few to explain the observed dispersion of the IR colors within the TC. In order to get quantitative measurements of the abundances variations, we compute the best fit of the parameters ($\chi, Y_{PAH}, Y_{VSG}$) in Table \ref{tab:mod_param_3f}. Even though the uncertainties on the radiation field intensity might be significantly higher than those on the abundances, we also give the best-fit values of $\chi$ in Table \ref{tab:mod_param_3f}. Relative to the TC reference, both the PAHs and VSGs abundances are lowered by a factor $\sim$ 3-4 in filament \#3 and increased by a factor $\sim 2-5$ in filament \#2 and \#6. The required increase is smaller within filament \#4 (a factor $\sim$ 1.2-2). In Figure \ref{fig:60mdl100}, variations in the IRIS100/MIPS160 color corresponds to $\chi/\chi_0 \sim 1$ for filaments \#1 and \#3 and $\chi/\chi_0 \sim 3$ for filaments \#2 and \#6. If this interpretation is right, we observe that the most illuminated filaments are those with the largest proportion of PAHs and VSGs. Reciprocally, the less illuminated filaments are those with the smallest contribution of PAHs and VSGs. An alternative possibility -- that would need to be tested with further modelling -- is that variations in the IRIS100/MIPS160 color trace changes in the VSGs size distribution. We also note that the variations of the PAHs and VSGs abundances are not equal. Within the uncertainties, filaments \#1, \#3 and \#6 exhibit the same relative variations for both PAHs and VSGs abundances, while filaments \#2 and \#4 require a relative increase of the PAHs abundance by 3 and 1.5 times that of the VSGs abundance, respectively.

These abundances variations occur on sub-parsec distances as the characteristic width of the filaments within the ``3 fingers" sub-region is about 20\arcmin\ or 0.8 pc (see Fig.~\ref{fig:3f_cuts}). They also occur within translucent sections of the cloud as the visual extinction of the filaments is about a few magnitudes at most (see Table \ref{tab:3f_cuts_colors}). Since the filaments were selected in an unbiased way, except for avoiding the zodiacal light, we could infer that similar dust size distribution variations take place also through the entire TC.

\subsection{Evolutionary processes}

The dust size distribution variations observed within the TC can be placed into a broader context. For instance, \citet{Boulanger1990} and \citet{Bernard1993} observed enhanced abundances of the smallest dust particles at cloud surfaces. \citet{Miville2002} detected an increase of the PAH abundance within a specific H~I velocity component in Ursa Major. However, these previous studies, based on IRAS and ISOCAM observations, lacked the sensitivity and field of view of the recent \textit{Spitzer} observations. The IRAC 8 $\mu$m and MIPS 24 $\mu$m data provide much higher spatial information on the PAHs and VSGs emission than in past studies. On the one hand, they reveal new structures within the cloud that remain unnoticed either in the $A_V$ or MIPS 160 $\mu$m map. On the other hand, within the ``clean" ecliptic latitude strip that we use for this analysis, there is no IRAC 8 $\mu$m and MIPS 24 $\mu$m counterpart to the bright structures in the $A_V$ and MIPS 160 $\mu$m images.

We relate the observed abundances variations to mass exchange between the small dust particles and the biggest grains. In the average model of the TC, the PAHs and VSGs contribute to 9\% of the total dust mass. Within filaments \#3 and \#6, which represent the two extreme cases of our analysis, the contribution of PAHs and VSGs to the total mass of dust ranges from 3 to 27\%. The mass fraction of PAHs and VSGs is 12\% in the original model of \citet{Desert1990}. We compute the equivalent fraction of carbon mass that is involved in those exchanges. We apply a carbon-silicate dust mass proportion of 37:63 which fits the diffuse ISM extinction curve \citep{Draine1984,Weingartner2001c}. We use a carbon abundance $[C/H]_{gas} = 161 \pm 17$ and $[C/H]_{total} \sim 330$ C atoms per million H determined along translucent lines of sight, even though these values are still subject to debate \citep[][and references therein]{Sofia2004}. For the TC, the average fraction of carbon within the smallest dust particles (PAHs and VSGs) is about 26\% relative to the total mass of carbon in dust particles and 13\% relative to the total mass of carbon in both dust and gas. Within filament \#3, where the relative abundance of PAHs and VSGs is the lowest, these figures are about 8\% and 4\% respectively. Within filament \#6, where the contribution of PAHs and VSGs is the largest, the values are about 72\% and 37\% respectively. Therefore, a significant fraction of the dust and carbon mass moves in and out of the narrow size range of stochastically heated PAHs and VSGs. These dust/carbon mass exchanges occur at the cloud surface, for moderate values of the visual extinction, well below the threshold where ice formation occurs \citep[e.g. extinction thresholds of $4.3\pm1.0$ mag for $\rm{CO_2}$, $3.2\pm0.1$ mag for $\rm{H_2O}$ and $6.7\pm1.6$ mag for CO in the Taurus complex,][]{Whittet2003,Whittet2007}.

\citet{Mathis1994} and \citet{Whittet2004} suggested that changes in the UV extinction curve could be accounted for by the formation of hydrocarbon coatings on grains small enough (size $<$ 20 nm) to contribute to the 2200 $\mbox{\AA}$ bump and far-UV rise of the extinction curve. Such a process could also account for the low abundance of PAHs and VSGs within the 160 $\mu$m  bright regions of the TC. The timescale for coagulation depends not only on the density but also on the relative velocities between grains. Grain relative velocities are set by turbulent motions and depend on the grain size. \citet{Yan2004} have calculated the relative grain motions arising from magnetohydrodynamic turbulence as a function of ISM physical conditions. They find that the largest velocities are always those of BGs. In that case, small grains are expected to condense on large grains faster than they mutually coagulate \citep{Draine1985}. However, small grains can also be efficiently decoupled from gas motions by large velocity gradients at small scales, known to exist in turbulence and not included in the \citet{Yan2004} work \citep{Falgarone1995,Pety2003}. If small grains acquire relative velocities comparable to those of BGs, PAHs and VSGs coagulation would occur faster on small grains than on BGs alone, because the total surface in small grains is larger.

Shattering of hydrocarbon grain mantles in grain-grain collisions is a likely process to explain the local enhancements of PAHs and VSGs abundances \citep{Serra2008}. Shattering occurs above some collision energy threshold \citep{Jones1996}. Here, turbulence may play a signiÞcant role in creating the necessary relative velocities between grains. The correlation between dust emission at 8 $\mu$m and 24 $\mu$m and specific structures in the gas velocity thus appears like a promising perspective for a future study. Once specific measurements of velocity gradients are available, the spatial scales over which the observed variations occur could be converted into their equivalent timescales.

\section{Conclusions}

We have presented Spitzer images of the Taurus Complex. We have combined these images with IRAS and DIRBE observations to characterize the diffuse emission across the cloud from mid to far-IR wavelengths. The sensitivity, spatial resolution and wavelength coverage of the observations permits analysis of small spatial scale variations of the dust emission. The data interpretation highlights evidence of dust evolution occurring within the translucent sections of an archetype example of quiescent molecular clouds.

The comparison between IRAS and MIPS observations at 60, 70, 100 and 160 $\mu$m reveals a strong spatial correlation between wavelengths although we can identify some variations. The average colors of the TC provide us with a FIR spectral energy distribution that is steeper than that of the Galactic diffuse medium. This difference suggests colder BGs. The average dust temperature is about 14.5K with a dispersion of $\pm 1$K across the cloud. The FIR dust opacity is tightly correlated with the extinction derived from 2MASS stellar colors. The FIR opacity to extinction ratio we infer from this correlation is a factor 2.2 larger than the average value for the diffuse ISM. This increase in opacity and the attenuation of the radiation field both contribute to the lower emission temperature of the large grains. 

The structure of the TC diffuse emission significantly changes at MIR wavelengths. Due to zodiacal light contamination, the \textit{Spitzer} maps only offer a partial view at the MIR emission from stochastically heated PAHs and VSGs. Outside the zodiacal bands, where the contamination is small, the 8 and 24 $\mu$m maps are similar but there are no 8 and 24 $\mu$m counterparts to the brightest 160 $\mu$m emission features.  The 8 and 24 $\mu$m images reveal filamentary structure that is strikingly inconspicuous in the 160 $\mu$m and extinction maps. The analysis of these filamentary structures shows significant variations within the MIR to FIR colors, from one filament to another and within one filament, over sub-parsec distances. 

We present model calculations that quantify the dependence of MIR to FIR colors on the radiation field intensity and the abundance of stochastically heated particles. To match the range in observed colors, we have to invoke variations by a factor of a few of both the interstellar radiation field intensity and the abundance of PAHs and VSGs. 

Our results should motivate theoretical studies of the evolutionary  processes, grain shattering and coagulation, which could account for the observed changes in the spectral energy distribution of the dust emission. The correlation between dust emission within IRAC 8 $\mu$m and MIPS 24 $\mu$m and specific structures in the gas velocity would allow determination of the potential role of turbulence in driving changes in the abundance of the smallest dust particles.

\acknowledgements
The authors would like to thank Laurent Cambresy for providing the 2MASS extinction map. They also thank Paul Goldsmith, Vincent Guillet and the anonymous referee for their questions and comments, which have significantly improved the paper.

This work is based in part on observations made with the Spitzer Space Telescope, which is operated by the Jet Propulsion Laboratory, California Institute of Technology under a contract with NASA.

{\it Facilities:} \facility{\textit{Spitzer}}.

\bibliographystyle{aa}
\bibliography{taurus}

\clearpage

\begin{table}[!b]
\caption{Average DIRBE colors and DIRBE 100 $\mu$m to IRIS 100 $\mu$m ratio across the TC}
\begin{center}
\begin{tabular}{cc}
\tableline
\tableline
Ratio & Value \\
\tableline
DIRBE60/DIRBE100 & $0.17\pm0.02$ \\
DIRBE100/DIRBE140 & $0.29\pm0.02$ \\
DIRBE140/DIRBE240 & $0.81\pm0.05$ \\
\tableline
DIRBE100/IRIS100 & $0.82\pm0.01$ \\
\tableline
\end{tabular}
\end{center}
\label{tab:dirbe}
\end{table}

\begin{table}[h]
\caption{MIR-to-FIR colors of the "three fingers" region and the whole TC.}
\begin{center}
\begin{tabular}{ccc}
\tableline
\tableline
Color	& \multicolumn{2}{c}{Value} \\
		& "Three fingers"	& TC \\ 
\tableline
$IRIS60/MIPS70$	& $0.63\pm0.22$	& $0.46\pm0.07$ \\
$IRIS60/IRIS100$	& $0.18\pm0.02$	& $0.20\pm0.02$\\
$MIPS70/IRIS100$	& $0.31\pm0.15$	& $0.35\pm0.05$ \\
$IRIS100/MIPS160$	& $0.27\pm0.01$	& $0.25\pm0.01$ \\
\tableline
\end{tabular}
\end{center}
\label{tab:3f_colors}
\end{table}

\clearpage

\begin{table}[h]
\caption{Dust components abundances and interstellar radiation field intensity used within our model for the best fit of the filaments.\label{tab:mod_param_3f}}
\begin{center}
\begin{tabular}{c c c c c c c}
\tableline
\tableline
Filament			& \#1	& \#2	& \#3	& \#4	& \#5	& \#6	\\
\tableline
$Y_{VSG}/Y_{VSG,0}$	& $0.92\pm0.05$	& $1.8\pm0.2$	& $0.25\pm0.04$	& $1.3\pm0.1$	& $0.48\pm0.25$	& $3.3\pm0.1$ \\
$Y_{PAH}/Y_{PAH,0}$	& $0.92\pm0.14$	& $5.2\pm0.8$	& $0.33\pm0.04$	& $1.9\pm0.3$	& $3.3\pm0.5$	& $3.8\pm0.6$ \\
$\chi/\chi_{0}$	& 0.9	& 2.5 	& 0.8	& 1.1	& 1.1	& 3.2 \\
\tableline
\end{tabular}
\end{center}
\end{table}

\clearpage

\begin{table}[h]
\caption{MIR-to-FIR colors of the filaments within the "three fingers" region, normalized to MIPS 160 $\mu$m, in $\times10^{-3}$. The background subtracted peak flux within MIPS 160 $\mu$m channel as well as the equivalent visual extinction are also given.}
\begin{center}
\begin{tabular}{c r r r r r r}
\tableline
\tableline
Filament			& \#1	& \#2	& \#3	& \#4	& \#5	& \#6	\\
\tableline
IRAC8/MIPS160	& 9.1	& 49		& 3.4	& 18		& 34		& 35		\\
MIPS24/MIPS160	& 8.7	& 32		& 2.6	& 14		& 8.5	& 32		\\
IRIS60/MIPS160 	& 51		& 77		& 30		& 58		& 82		& 120	\\
IRIS100/MIPS160 	& 290	& 490	& 270	& 320	& 310	& 500	\\
\tableline
MIPS160 (MJy/sr)	& 37		& 6		& 66		& 12		& 4		& 9		\\
$A_V$ (mag)		& 2.1	& 0.3	& 3.8	& 0.7	& 0.2	& 0.5	\\
\tableline
\end{tabular}
\end{center}
\label{tab:3f_cuts_colors}
\end{table}

\end{document}